\documentstyle[12pt]{article}

\def\NPB#1#2#3{Nucl. Phys. B{#1} (19#2) #3}
\def\PLB#1#2#3{Phys. Lett. B{#1} (19#2) #3}

\def\PRL#1#2#3{Phys. Rev. Lett. {#1} (19#2) #3}

\def\MODA#1#2#3{Mod. Phys. Lett.  {#1} (19#2) #3}

\def\yzero{\smash{\hbox{$y\kern-4pt\raise1pt\hbox{${}^\circ$}$}}}

\def\b{\beta}

\def\-{\hphantom{-}}

\def\s2{\frac{1}{\sqrt2}}
\def\s22{\frac{1}{2\sqrt2}}

\def\beq{\begin{equation}}
\def\eeq{\end{equation}}
\def\beqa{\begin{eqnarray}}
\def\eeqa{\end{eqnarray}}

\def\IF{\relax{\rm I\kern-.18em F}}
\def\II{\relax{\rm I\kern-.18em I}}
\def\IP{\relax{\rm I\kern-.18em P}}
\def\IR{\relax{\rm I\kern-.18em R}}
\def\inbar{\vrule height1.5ex width.4pt depth0pt}
\def\IC{\relax\hbox{\kern.25em$\inbar\kern-.3em{\rm C}$}}

\def\Dsl{\,\raise.15ex\hbox{/}\mkern-13.5mu D} 
\def\IZ{Z\kern-.3em  Z}
\def\cp#1{\relax\ifmmode {\IP\kern-2pt{}_{#1}}\else $\IP\kern-2pt{}_{#1}$\fi}

\begin{document}

\makeatletter
\@addtoreset{equation}{section}
\makeatother
\renewcommand{\theequation}{\thesection.\arabic{equation}}
\pagestyle{empty}
\rightline{IASSNS-HEP-98/8}
\rightline{hep-th/9803054}
\begin{center}
\huge{Towards mass deformed \\
$N=4$ $SO(n)$ and $Sp(k)$ theories\\
from brane configurations \\}
\bigskip
\large{Angel M. Uranga \footnote{e-mail: uranga@ias.edu \\
Research supported by the Ram\'on Areces Foundation}\\[2mm]}
\large{\em School of Natural Sciences, Institute for Advanced
Study\\[-0.3em]
Olden Lane, Princeton NJ 08540\\[1mm]}

\bigskip

\bigskip

{\small {\bf Abstract} \\[5mm]}
\end{center}

\begin{center}
\begin{minipage}[h]{14.0cm}

{\small
We study the introduction of orientifold six-planes in the type
IIA brane configurations known as elliptic models. The $N=4$ $SO(n)$ and
$Sp(k)$ theories softly broken to $N=2$ through a mass for the adjoint
hypermultiplet can be realized in this framework in the presence of two
orientifold planes with opposite RR charge. A large class of $\b=0$
models is solved for vanishing sum of hypermultiplet masses by embedding
the type IIA configuration into M-theory. We also find a geometric
interpretation of Montonen-Olive duality based on the properties of the
curves. We make a proposal for the introduction of non-vanishing sum of
hypermultiplet masses in a sub-class of models.

In the presence of two negatively charged orientifold planes and four
D6-branes other interesting $\beta=0$ theories are constructed, e.g.
$Sp(k)$ with four flavours and a massive antisymmetric hypermultiplet.
We comment on the difficulties in obtaining the curves within our
framework due to the arbitrary positions of the D6-branes.}

\end{minipage}
\end{center}
\newpage

\setcounter{page}{1}
\pagestyle{plain}
\renewcommand{\thefootnote}{\arabic{footnote}}
\setcounter{footnote}{0}

\section{Introduction}

The study of the low-energy dynamics of branes has become a powerful tool
to understand old and new phenomena in supersymmetric gauge field
theories (for a recent review, see \cite{giveon}).

For theories with eight supercharges the basic setup is provided
by the configurations of type IIB NS5-branes, D3-branes and D5-branes
introduced in \cite{hw}, where they were employed in the study of
several
phenomena in three dimensions. By performing T-dualities along appropriate
directions, many interesting results have been obtained for theories in
several dimensions.

To our main interest in this paper is the fact that the type IIA
configurations leading to four-dimensional $N=2$ theories provide, when
embedded in M-theory, a fairly systematic construction of the
corresponding Seiberg-Witten curves \cite{swsu2}, which govern the
dynamics of the theories on the Coulomb branch. These curves were known
from the gauge field theory viewpoint mainly in
the examples in which the hyperelliptic ansatz gives the
correct answer, i.e. essentially for classical simple groups with
fundamental flavours.

The techniques developed in \cite{witten4d},
however, allow to an easy rederivation of all the results concerning
unitary gauge groups \cite{swsu2,puresun,hanoz,aps}, and an
authomatic
generalization to product group theories with matter in bi-fundamentals
and
fundamental flavours.

An analogous success has been obtained in reproducing the curves for
orthogonal \cite{danielsson, landsteiner,hanany,argshap} and
symplectic
groups \cite{argshap,dhkp} by the introduction of orientifold
four-planes \cite{lll,sospbrane} (based on previous considerations made
in \cite{johnson}), and orientifold six-planes \cite{landlop} again with
generalizations to product group theories. In this last reference new
curves for $SU(n)$ gauge theories with matter in the symmetric or
antisymmetric representations were also proposed.

Another interesting family of models is that with $SU(n)$ gauge group and
a hypermultiplet in the adjoint. The case $n=2$ was studied in
\cite{swsu2}, and the general case was solved from the field theory
point of view in \cite{dw}. When the adjoint mass vanishes the theory has
enhanced
$N=4$ supersymmetry, and the SW curve is rather trivial, a set of
identical tori giving the microscopic $\tau$ parameter of the gauge
theory. When the adjoint mass is not zero, the supersymmetry is softly
broken to $N=2$ and non-trivial dependences on moduli space appear. The
description in \cite{dw} was successfully recovered {\em via} brane
configurations as a particular case of
the elliptic models in \cite{witten4d}, in which one of the directions of
the spacetime in which the branes are embedded is made compact (more
concretely, $x^6$, along which the D4-branes are finite). Again, a
natural generalization to product group theories arose as well.

\medskip

The purpose of this article is to discuss the construction of the SW
curves for the mass deformed $N=4$ theories with symplectic and orthogonal
gauge groups, by means of the brane configurations. The natural framework,
as we will argue, is the introduction
of orientifold six-planes in these elliptic models, a generalization not
studied so far. Many other new theories can be studied analogously, some
of them with simple groups (e.g. $SU(n)$ with symmetric {\em and}
antisymmetric hypermultiplets), some with product gauge groups. They are
however interesting because provide further examples of finite gauge
theories, a feature that arises naturally in the context of elliptic
models.

The plan of the paper is as follows. We start with a review of the
elliptic models in \cite{witten4d} in Section 2, in order to set up the
notation and stress some facts for future use.

In Section 3 we introduce
the classical type IIA configurations with the inclusion of the orientifold
six-planes. An interesting feature of having compactified the $x^6$
direction is that the configurations contain {\em two} orientifold planes.
Therefore two large classes of models can be discussed, which differ, to
put it briefly, in the relative sign of the orientifold charges.
We describe the type of theories that can be obtained in this
framework, stressing that in particular cases they reduce to theories with
adjoint hypermultiplets, or other interesting theories with vanishing $\b$
function.

In Section 4 we introduce several ingredients required to embed the
brane configuration in M-theory. We comment on two inmediate difficulties
encountered. We analyze them in the simpler case of non-compact $x^6$,
which can be thought of providing a local description of the elliptic
models. Both difficulties are solved by a brane creation
effect due to the RR charge of the orientifolds (and D6-branes when
present). Two other interesting geometrical facts about the curves of
theories in the presence of negatively charged orientifolds are also
uncovered.

In Section 5 we lift a large class of models to M-theory and construct
their SW curves. As in reference \cite{witten4d} we start with the case in
which the sum of the masses of the hypermultiplets vanishes. The use of
the results of Section 4 makes it possible to construct them in close analogy
with the elliptic models in \cite{witten4d}. We illustrate the consistency
of our description by showing how the proposed curves reproduce flat
directions connecting the different models. We also compare the curves
with known results in some simple cases where these latter exist.
Finally, we propose a geometric interpretation of Montonen-Olive duality
for orthogonal and symplectic groups, as well as its generalization to
product group theories.

A proposal for the introduction of non-vanishing sum of hypermultiplet
masses is presented in Section~6. In particular, we present a candidate
curve for the $Sp(k)$ theories with an adjoint hypermultiplet with
arbitrary mass.

We finish with several comments and conclusions in Section~7.

\section{Elliptic models}

In this section we review the brane configuration of \cite{witten4d}
that
provides the solution for the $N=2$ $SU(k)$ gauge theory with one adjoint
hypermultiplet, as well as the generalization to product group theories.
While the contents of this section are not original, we stress a few
points which will turn out to be useful in future sections.

The basic type IIA setup is as follows: The NS5-branes are extended in the
$x^0$, $x^1$, $x^2$, $x^3$, $x^4$ and $x^5$ directions, being located at
equal values in $x^7$, $x^8$ and $x^9$ and differing in their $x^6$
positions. The D4-branes span the $x^0$, $x^1$, $x^2$, $x^3$ and $x^6$
directions, being finite in the last one, in which they are suspended
between NS5-branes. The $x^6$ direction is taken compact, i.e. a $S^1$ of
length $2\pi L$, so the configuration can be equivalently described as an
infinite periodic array on the covering spacetime.

In principle it would be possible to include D6-branes in the
construction, without breaking further supersymmetries, but the
requirement of having zero or negative \mbox{$\beta$-function} for all the
factors in the gauge
theory on the D4-branes excludes such possibility.

This condition also
implies that the number of D4-branes suspended between a pair
of neighbouring NS5-branes is the same for all such pairs, $k$. On the
four-dimensional noncompact part of the D4-brane world-volume, there is a
$N=2$ $SU(k)^n\times U(1)$ gauge theory, where $n$ is the number of
NS5-branes. The hypermultiplet content is
given by bi-fundamentals with respect to neighbouring factors (the
last $SU(k)$ being neighbour of the first, due to compactness in
the $x^6$ direction), in a quiver-like fashion. The $\beta$-function is
zero for all factors.

Naively, the sum of the masses of the hypermultiplets
vanishes, but as will become clear by the end of this section, this
feature can be avoided by an appropriate choice of the M-theory spacetime
(or the turning on of certain backgrounds in the IIA language). Such mass
parameter will be referred to as a `global mass' in what follows.

When $n=1$, the gauge theory reduces to $U(k)$ with an adjoint
hypermultiplet. It has $N=4$ supersymmetry if the global mass (adjoint
mass) vanishes, and is softly broken to $N=2$ if
such global mass term is introduced.

Many field theory phenomena can be understood geometrically
in the brane picture. To our main interest,
the motion of the D4-branes in the $x^4$, $x^5$ directions probes the
Coulomb branch of the gauge theories just described.

\medskip

The absence of D6-branes can also be derived by noticing that they
constitute magnetic sources for the Ramond-Ramond $U(1)$. The
corresponding flux could in principle escape to infinity on the noncompact
directions transverse to the D6-branes. However, since $x^6$ is now
compact, there are only two of them, namely $x^4$ and $x^5$ (which we
subsequently arrange in the complex variable $v$), and the gauge
potential would
be logarithmically divergent at long distances. This behaviour simply
indicates the running of the coupling constant with $v$, so the
$\beta$-function does not vanish. There is a one-to-one correspondence
between brane configurations with zero net RR charge and gauge
theories with vanishing $\beta$-functions. This fact will be a key
ingredient in the discussion in Section 3.

\medskip

Quantum effects on the Coulomb branch can be studied by embedding the above
configuration in M-theory, by introducing the new compact
dimension $x^{10}$. The part of spacetime in which brane motion takes
place has the structure $E_{\tau}\times \IC_v$, where $\IC_v$ is a complex
plane parametrized by $v$, and $E_{\tau}$ is a two-torus
(hence the name `elliptic' for these models) parametrized by $x^6$,
$x^{10}$. Its complex structure parameter determines the microscopic
coupling for the gauge theory \footnote{If the gauge theory contains
several factors, $\tau$ gives the coupling of the diagonal subgroup, the
individual couplings being determined by the NS5-brane positions in
$E_{\tau}$.}.

In this regime, the D4-branes gain a new dimension, and wrap
$E_{\tau}$, and are located at a point in $v$. The NS5-branes maintain
their dimensionality, and span $v$, being located a points in $E_{\tau}$.
Actually, the whole brane configuration becomes a single smooth M5-brane,
whose worldvolume is a Riemann surface, $\Sigma$. This is
the Seibeg-Witten curve for the theory, i.e. the periods
of a suitable differential form on this curve determine the exact
non-perturbative prepotential for the gauge theory. The differential form
is found to be $v\, dx/y$ by analyzing the masses of BPS saturated states,
which are realized as membrane states with boundaries on the M5-brane.
\cite{klmvw}

The curve can be
described as a $k$-fold cover of $E_{\tau}$, since it reproduces the $k$
D4-branes in the type IIA limit, each of which wraps $E_{\tau}$ once. It
is
described by an equation of the form
\beq
v^k\, +\, f_1(x,y)\, v^{k-1}\, +\, \ldots\, +\, f_k(x,y)\, =\, 0,
\label{witcover}
\eeq
with $f_l(x,y)$ being meromorphic functions on $E_{\tau}$, thus rational
in the
variables $x$, $y$, which parametrize the elliptic curve when written in
the Weierstrass form
\beq
y^2\, =\, (x-e_1(\tau))\, (x-e_2(\tau))\, (x-e_3(\tau)).
\label{weierstrass}
\eeq

These functions must have, at most, simple poles at the locations of
NS5-branes, and are holomorphic at the point at infinity (unless a
NS5-brane is placed there, in which case they have a simple pole at
infinity).

A check of this description comes from matching the counting of
parameters in the equations with that in the gauge theory. An elliptic
function with $n$ poles depends on $n$ parameters, so the curve $\Sigma$
we have constructed depends on $kn$ parameters. They correspond to
$n(k-1)+1$ moduli of the Coulomb branch of the field theory, plus $n-1$
bare masses for the bi-fundamentals (we have already mentioned the
vanishing of the sum of masses). Altogether these determine the
positions for
D4-branes. It is also possible to identify the independent masses as the
residues of the differential form $f_1(x,y)\,dx/y$ at $n-1$ poles.
As already mentioned in a footnote, the gauge couplings are encoded in
$\tau$ and the positions of the NS5-branes in $E_{\tau}$.

Let us pause for a moment to provide a more explicit description of the
$n$ parameters
each of the $f_i(x,y)$ depends on. This analysis will turn out to be
useful in the generalizations we study in Section 5. Any elliptic function
has an equal number of zeros and poles. Moreover, there is a global
restriction on their locations
\footnote{Since it makes use of the addition law on
a cubic curve, this restriction is most easily stated in terms of the
uniformizing
complex variable $z$, which parametrizes the torus as the quotient of
$\IC$ by translations in a bi-dimensional lattice of periods
$\Lambda$. The mentioned addition law becomes usual addition in $\IC$. If
we denote the positions of the zeroes by $z\, =\, a_1,\ldots , a_n$ and
those of poles by $z\,=\, b_1,\ldots ,b_n$, the mentioned restriction
reads
\beq
(a_1\, + \ldots +\, a_n) \, -\, (b_1\, +\ldots +\, b_n)\, =\, \Omega ,
\label{diffsum}
\eeq
with $\Omega \in \Lambda$ a period. The elliptic function can
be very explicitly constructed in terms of the Weierstrass
$\sigma$-function (see \cite{chandra} for details), as follows:
\beq
f(z)\, =\, C\, \frac{\sigma(z-a_1)\ldots
\sigma(z-a_n)}{\sigma(z-b_1)\ldots \sigma(z-b_n-\Omega)},
\label{sigmas}
\eeq
where $\sigma(z)$ is an entire (not elliptic) function with simple zeroes
at the points
in $\Lambda$.}.
Since in our case the positions of the poles are fixed, there are $n-1$
parameters controlling the positions of the zeroes (the $n$-th being fixed
by  eq.(\ref{diffsum}). The remaining parameter is simply the overall
multiplicative constant $C$ in eq.(\ref{sigmas}).

This description of the parameters is well-suited e.g. for the discussion
of Higgs branches. When the mass of a bi-fundamental vanishes there is a
flat
direction in the gauge theory along which this hypermultiplet gets a vev
and breaks the gauge symmetry
\beqa
SU(k) \times SU(k) & \rightarrow & SU(k)_D \nonumber\\
({\bf k},{\bf k}) \quad\quad & & {\bf 1}
\label{Higgs1}
\eeqa
the final $SU(k)$ being the diagonal subgroup of the two initial ones.
From the original bi-fundamental, $k^2-1$ modes are eaten by the Higgs
mechanism and only a singlet remains. Note that the final $SU(k)$ has
bi-fundamentals with respect to its neighbours in the chain (even though
we have not displayed them in eq.(\ref{Higgs1})), so after the breaking
the theory is of the
same kind, but with one factor less. The number of parameters that
disappear is $k$, namely $k-1$ order parameters and one mass. It is easy
to conclude this Higgssing is reproduced in the brane picture by
recombining the D4-branes on both sides of a NS5-brane and removing
this last in the $x^7$, $x^8$, $x^9$ directions.

In the description provided by equation (\ref{witcover}) the adequate
procedure is to force one of the zeroes of each of the functions
$f_i(x,y)$ to match
the position of the pole corresponding to the NS5-brane to be removed.
This requires the
tuning of $k$ parameters, and yields a model of the same type but one
pole less, i.e. the NS5-brane has been removed from the configuration.

\bigskip

Let us end this section by reviewing the introduction of a global mass for
the hypermultiplets.
In order to do so, one needs, in
the type IIA picture, the D4-branes to come back to a shifted value of $v$
when they go around $x^6$. This is implemented by allowing the spacetime
to have the structure of a non-trivial affine $\IC_v$-bundle over
$E_{\tau}$,
instead of being the globally trivial product $\IC_v \times E_{\tau}$. The
curve $\Sigma$ is described patchwise. At any point but
infinity in $E_{\tau}$, expression (\ref{witcover}) remains valid, only
the behaviour
of the functions $f_i(x,y)$ at infinity is different. Instead of being
holomorphic, $f_i(x,y)$ has a pole of order at most $i$ at infinity (the
condition of having simple poles at the locations of NS5-branes is
unchanged). This introduces $k(k+1)/2$ additional parameters in the
equations, but they are determined by a good `glueing' with the
description at the point
at infinity, as follows. The adequate variables at a neighbourhood of this
point are $u$, $z$ and ${\tilde v}$, related to the
previous ones by
\beqa
x \; = \; u^{-2} \quad ; \quad y \; = \; z\, u^{-3} \quad ; \quad
{\tilde v} \; = \; v+\frac{m}{2k} \frac{y}{x}
\label{patchmass}
\eeqa

The equations describing the curve $\Sigma$ are obtained from
(\ref{weierstrass}) and (\ref{witcover}) by simple substitution. The
spectral cover can be recast as
\beqa
{\tilde v}^k\, +\, {\tilde f}_1(z,u)\, {\tilde v}^{k-1}\, +\, \ldots\, +\,
{\tilde f}_k(z,u)\, =\, 0,
\label{covertilde}
\eeqa
where the ${\tilde f}_i(z,u)$ is also an elliptic function, which differs
from $f_i$'s, and in
principle has a poles of order at most $i$ at $u=0$.
Since there are no NS5-branes at infinity, there should not be any
pole at $u=0$ in this equation \footnote{If one NS5-brane is located at
infinity the condition is that at most a simple pole should be
left.}. So we force a zero of order $i$ in the function ${\tilde
f}_i(z,u)$ to cancel the pole. These are $k(k+1)/2$ conditions that can be
used to eliminate extra
parameters that had been introduced in the equations. The parameter
$m$ appears as the residue of a simple pole of $f_1(x,y)$ at infinity, and
provides the global mass sought for.

In \cite{dw} a very explicit description of the functions $f_i(x,y)$
was presented for the case of having one NS5-brane, $n=1$. We will not
present
this more detailed treatment here, even though we will use analogous
ideas in Section~6.

\section{Type IIA configurations with orientifold six-planes}

The inclusion of orientifold planes in brane configurations is an
efficient and natural way to study field theories with orthogonal and
symplectic gauge groups. In the type IIA configurations in
\cite{witten4d} it is possible to introduce orientifold 4-planes  or
orientifold 6-planes (parallel to the D-branes of equal dimensionality)
while preserving $N=2$ supersymmetry in the
four-dimensional theory.

The effect of O4-planes has been analyzed in this context in
\cite{johnson,lll, sospbrane} for the case in which $x^6$ is
non-compact. In \cite{lll} the case of compact $x^6$ was also studied,
but it is easy to conclude that the mass deformed $N=4$
theories cannot be obtained in this framework. For example, the
gauge groups obtained are necessarily semisimple, products of
orthogonal and symplectic groups, with an {\em even} number of factors.
Also the $Z_2$ projection imposed by the orientifold eliminates the
possibility of turning on hypermultiplet masses.

\smallskip

The inclusion of O6-planes has been analysed in \cite{landlop} for the
case in which $x^6$ is non-compact. The purpose of the present article is
to understand the corresponding elliptic models. This extension is not as
straightforward as may look at first sight, and some subtleties in the
construction of the curves will be presented (and solved) in Section 4.
For the moment we start by discussing in the type IIA limit the
family of theories which can be
obtained through these elliptic models, in order to motivate further
investigation beyond the classical limit. We will show that indeed the
$SO(n)$ and $Sp(k)$
mass deformed $N=4$ theories, as well as other interesting finite
theories, can be thus constructed.

\medskip

The type IIA setup is exactly that introduced in Section 2, save for the
orientifold planes introduced by the $Z_2$ quotient.
This $Z_2$ acts as a simultaneous inversion of the coordinates $x^6$ and
$v$, along with the operation $(-1)^{F_L}$ and orientation reversal on the
type IIA string world-sheet. In the quotient, the compact $x^6$ direction
is mapped to an interval, whose boundary corresponds to two
orientifold six-planes fixed under the $Z_2$ action. They extend along the
directions $x^0$, $x^1$,
$x^2$, $x^3$, $x^7$, $x^8$, $x^9$, and are located at the fixed points
in $v=0$ and $x^6=0,\pi L$. We will usually work
on the double cover, and consider the $x^6$ dimension as a $S^1$ where the
dynamics is symmetric under the $Z_2$.

There are two kinds of O6-planes, which differ in the sign of their
RR charge, and in the projections they induce on D6- and D4-branes. The
first kind has charge $-4$ (in D6-brane charge units) on the double
cover.
We will denote it by $O6^-$. It produces a gauge symmetry $SO(2N)$ when
$N$ D6-branes (along with their mirrors) approach it \footnote{These
orientifold planes can have a D6-brane stuck on them, in which case the
gauge symmetry is $SO(2N+1)$.}. On the D4-branes, however, the
projection acts differently and the gauge group obtained when $M$
D4-branes
(plus mirrors) approach it is $Sp(M)$ \footnote{Our convention is that
the fundamental representation of $Sp(M)$ has dimension $2M$.}. This is
consistent with the fact that the gauge group on the D6-branes is the
global symmetry group of the gauge theory on the D4-branes. The second
kind of O6-plane has charge $+4$ (we denote it by $O6^+$), and projects
onto
symplectic groups on D6-branes and orthogonal groups on D4-branes. Even
though in the perturbative description the two kinds look very symmetric
one with respect to the other, their non-perturbative properties differ
markedly \cite{witfut,landlop}.

When constructing the brane configuration we can choose independently the
kind
of O6-plane we locate at each fixed point. However, let us recall that,
by the reasoning mentioned in Section 2, theories with zero
$\b$ function for all gauge factors can only be obtained when the net
magnetic charge vanishes. Thus one is led to reject the configuration with two
$O6^+$'s, since there is no way of cancelling their positive charge
(anti-D6-branes would break all the supersymmetries). On the other hand,
the mixed configuration with one orientifold plane of each kind has zero
net charge and can be directly considered for our purposes. The last
possibility, with both O6-planes of charge $-4$,
requires
the introduction of four D6-branes (plus their mirrors) in order to yield
theories with zero $\b$ function (the cases with negative $\b$
functions for some of the gauge factors can be obtained by sending some or
all the D6-branes to
infinity in $v$, which corresponds to decoupling some hypermultiplets by
making them infinitely massive). Configurations T-dual to these have
recently been considered in \cite{witfut, sixd}.

Next, one introduces a number of NS5-branes, and D4-branes suspended
between them,  with the only restriction that the configuration must be
symmetric under the $Z_2$, so that it makes sense in the quotient. As in
the elliptic models of Section 2, there are strong constraints in the
number of D4-branes allowed between a given pair of neighbouring
NS5-branes. A novelty
is that this number differs in the different factors in a semisimple
gauge group, but the patterns are completely determined by the
requirement of having $\b\leq 0$ for all factors. A proof that the
non-positiveness of the $\b$ functions implies the spectra below can be
carried out as a modified version of the corresponding argument in
\cite{witten4d}. However we do not present it here, and directly turn to
classifying
the type of theories which arise from both choices of orientifold
background.

\subsection{The $O6^+$-$O6^-$ background}

In the configuration with O6 planes of both kinds, the possibilities are:

{\bf i)} There are $p+1$ NS5-branes at generic values of $x^6$ (not
coinciding
with those of the O6-planes), and $p+1$ more at the symmetric positions.
Upon
the introduction of the D4-branes, the resulting gauge group and
hypermultiplet content are
\beqa
&Sp(k) \times SU(2k+2) \times \cdots \times SU(2k+2p) \times SO(2k+2p+2)
& \nonumber\\
&({\bf 2k},{\bf 2k+2},{\bf 1},\ldots,{\bf 1}) + \sum_{l=1}^{p-1} ({\bf
1},\ldots,{\bf 2k+2l},{\bf 2k+2l+2},\ldots,{\bf 1}) & \\
& + ({\bf 1},\ldots,{\bf 1},{\bf 2k+2p},{\bf 2k+2p+2})& \nonumber
\label{chorizo1}
\eeqa
The presence of the symplectic factor makes clear the only even numbers of
D4-branes are allowed to join the NS5-branes.
One easily checks that the $\b$ function vanishes for all factors. Notice
that the
gauge group on the D4-branes at values of $x^6$ near the $O6^-$ is
symplectic, while that of the D4-branes near the $O6^+$ is orthogonal. For
other factors, the gauge group is unitary because the $Z_2$ does not
restrict the motion of D4-branes within a factor, but relates it to a
`mirror factor', located elsewhere in $x^6$. This gauge theory is actually
a `compactified' version of theories considered in \cite{landlop}.
Also notice
that the U(1) factor that appeared in Section 2 is projected out by the $Z_2$
symmetry.

The simplest case in this family of models is $p=0$,  when there are two
NS5-branes at mirror positions. If we let $2k$
D4-branes join the NS5-branes on one side (that corresponding to the
$O6^-$), and $2k+2$ on the other ($O6^+$), the gauge group is
$Sp(k)\times SO(2k+2)$ and there
is one hypermultiplet in the bi-fundamental. This particular theory can be
alternatively constructed using elliptic models with O4-planes \cite{lll}.

\bigskip

{\bf ii)} We place $p+1$ NS5-branes plus their mirrors, and locate an
additional NS5-brane intersecting the $O6^-$ (so that the NS5-brane is
mapped to itself under the symmetry). In this case the gauge group that
arises from D4-branes near the $O6^-$ is unitary, and, in addition to
the bi-fundamental, this factor has a
hypermultiplet in the two-index antisymmetric representation. The numbers
of D4-branes suspended between pairs of NS5-branes are not constrained to
be even. The spectrum is:
\beqa
& SU(n) \times SU(n+2)  \times \cdots \times SU(n+2p) \times
  SO(n+2p+2) & \nonumber\\
& ({\bf \frac{n(n-1)}{2}},{\bf 1}, \ldots, {\bf 1}) +
  ({\bf n},{\bf n+2},{\bf 1},\ldots,{\bf 1}) + & \\
& + \sum_{l=1}^{p-1} ({\bf 1},\ldots,{\bf n+2l},{\bf n+2l+2},\ldots,{\bf
1}) + ({\bf 1},\ldots,{\bf 1},{\bf n+2p},{\bf n+2p+2}) & \nonumber
\label{chorizo2}
\eeqa
The $\b$ function vanishes for all factors.
Again, this is a `compactified' version of theories in \cite{landlop}.

The simplest and most interesting example in this family consists on
having just the NS5-brane that intersects the orientifold plane, and $n$
D4-branes going around the whole $x^6$ dimension. On the
other hand, it is also a rather peculiar construction, since orientifolds
of both signs seem to perform projections on the D4-branes. The
resulting theory has gauge group $SO(n)$ (due
to the $O6^+$ projection) and there is a hypermultiplet in the
antisymmetric
representation (due to the $O6^-$), which in fact is the adjoint. If its
mass
is zero the theory has $N=4$ supersymmetry. A very interesting question is
whether the brane configuration allows for the introduction of such
soft breaking mass,
or on the contrary the $Z_2$ symmetry forbids it. We will argue below that
such
deformation survives in the quotient and that the brane configuration
provides
a construction of the theory for arbitrary adjoint mass.

Notice that the obtention of the adjoint matter depends on having $x^6$
compact in an essential way, very much like in the $SU(k)$ theories of
Section~2. In this
respect, the interplay between both kinds of orientifolds to yield the
adequate projection on the gauge group and matter content is particularly
amusing.

\medskip

Let us finally comment on an interesting flat direction which connects
each model in this family to a model in class i). When $n$ is even, this
is easily understood as the Higgssing
\beqa
SU(2k) \;\;\; & \rightarrow & Sp(k) \nonumber\\
{\bf \frac{2k(2k-1)}{2}} & & \;\;\; {\bf 1}
\eeqa
triggered by a vev for the hypermultiplet in the antisymmetric
representation.
The gauge factors and matter hypermultiplets not shown do not take part in
the Higgssing. The
remaining singlet parametrizes the one-dimensional (in quaternionic units)
Higgs branch. In the brane picture, this corresponds to removing the
unpaired NS5-brane in the $x^7$, $x^8$, $x^9$ directions, and
recombining the D4-branes that were to its left and right.

When $n$ is odd the flat direction involves also {\em all} of the
bi-fundamentals, and the breaking, which is more complicated, proceeds as
\beqa
& SU(2k+1) \times SU(2k+3) \times \ldots \times SU(2k+2p+1)\times
SO(2k+2p+3) & \rightarrow \nonumber\\
\rightarrow & Sp(k)\times SU(2k+2)\times \ldots\times SU(2k+2p) \times
SO(2k+2p+2)
\label{comphig1}
\eeqa
The matter content, not shown for clarity, is that listed above for the
corresponding ii) and i) models, save for an extra singlet that remains
after the symmetry breaking, and which parametrizes a one-dimensional
Higgs branch. The brane interpretation of this process is as follows: One
must force one D4-brane from each gauge factor to lie at $v=0$, so it can
form a longer D4-brane which goes all the way around the $x^6$ circle
(notice how this lowers the rank of the `middle' unitary factors).
Then the unpaired NS5-brane is removed in $x^7$, $x^8$, $x^9$ along with
this D4-brane, so that only even numbers of D4-branes are left at each
interval in the brane configuration, as must be the case for a type i)
model. For this last step, of course, the D4-branes to the left and right
of the NS5-brane must be reconnected.

\bigskip

{\bf iii)} There is an analogous family of models in which the  unpaired
NS5-brane intersects the $O6^+$ instead of the $O6^-$. The resulting
gauge theory is
\beqa
& Sp(k) \times SU(2k+2)  \times \cdots \times SU(2k+2p) \times
SU(2k+2p+2) & \nonumber\\
& ({\bf 2k},{\bf 2k+2},{\bf 1},\ldots,{\bf 1}) + \sum_{l=1}^{p-1} ({\bf
1},\ldots,{\bf 2k+2l},{\bf 2k+2l+2},\ldots,{\bf 1}) + & \\
& + ({\bf 1},\ldots,{\bf 1},{\bf 2k+2p},{\bf 2k+2p+2}) + ({\bf
1},\ldots,{\bf 1},{\bf \frac{(2k+2p+2)(2k+2p+3)}{2}}) & \nonumber
\label{chorizo3}
\eeqa
In this case it is the factor near the $O6^+$ the one that becomes
unitary, and it has a hypermultiplet in
the two-index symmetric representation. One can easily check that
the $\beta$ functions of all factors vanish.

Again, the most interesting case arises when
only the unpaired NS5-brane is present. Letting $2k$
D4-branes be suspended, the theory has gauge group
$Sp(k)$ with matter in the symmetric, which is in fact the adjoint. So
this family provides the construction of $N=4$ $Sp(k)$ theories. Again,
it will be possible to introduce an arbitrary mass for the adjoint
hypermultiplet.

\medskip

In this case there is also a flat direction connecting these models to
those in the family i). It is easily understood as the breaking of
$SU(2k+2p+2) \to SO(2k+2p+2)$ by a vev for the hypermultiplet in the
symmetric representation. In the brane picture it amounts to removing the
unpaired NS5-brane, and recombining the corresponding D4-branes.

\bigskip

{\bf iv)} A last possibility is having two unpaired NS5-branes, located
each at a fixed point in the compact dimension. The gauge theory is
\beqa
& SU(n) \times SU(n+2)  \times \cdots \times SU(n+2p) \times
SU(n+2p+2)  & \nonumber \\
& ({\bf \frac{n(n-1)}{2}},{\bf 1}, \ldots, {\bf 1})
  + ({\bf 1},\ldots,{\bf 1},{\bf \frac{(n+2p+2)(n+2p+3)}{2}}) +
 ({\bf n},{\bf n+2},{\bf 1},\ldots,{\bf 1}) & \\
& + \sum_{l=1}^{p-1} ({\bf 1},\ldots,{\bf n+2l},{\bf
n+2l+2},{\bf 1},\ldots,{\bf 1})
  + ({\bf 1},\ldots,{\bf n+2p},{\bf n+2p+2})&  \nonumber
\label{chorizo4}
\eeqa
which is a mixture of cases ii) and iii), and again has $\beta=0$ for all
factors.

The simplest example in this family consists on
keeping only the NS5-branes intersecting the orientifold planes, and
considering $n$ D4-branes (and mirrors) joining them. The gauge
group is $SU(n)$ and there are matter hypermultiplets in the symmetric
and antisymmetric representations.

\medskip

Let us discuss the flat directions connecting these models to
other families. First, by using the hypermultiplet in the symmetric
representation the $SU(n+2p+2)$ factor can be Higgssed down to
$SO(n+2p+2)$, recovering models in class ii). As usual, in the brane
picture one simply removes the NS5-brane intersecting the $O6^+$ and
reconnects the corresponding D4-branes. Notice that in this process
there is not any essential difference between the even and odd $n$ cases.

In order to Higgs to models in the family iii), however, we must
distinguish these two cases. If $n$ is even, the flat direction involves
only the hypermultiplet in the antisymmetric representation. It performs
the breaking of $SU(2k)$ to $Sp(k)$. If $n$ is odd, a more complicated
Higgssing is required, rather analogous to (\ref{comphig1})
\beqa
& SU(2k+1)\times SU(2k+3)\times\ldots\times SU(2k+2p+1)\times
SU(2k+2p+3) & \rightarrow\nonumber\\
\rightarrow & Sp(k)\times SU(2k+2)\times\ldots\times SU(2k+2p)\times
SU(2k+2p+2)
\label{comphig2}
\eeqa
As in preceding cases, the brane interpretation consists on removing the
NS5-brane intersecting the $O6^-$ (along with a D4-brane at $v=0$ in the
odd $n$ case).


%

\subsection{The $O6^-$-$O6^-$ configuration}

The configurations in which both O6-planes are chosen with $-4$ charge,
and
four D6-branes (plus mirrors) are included, can be discussed analogously,
but the spectra are more cumbersome to list in the general case,
essentially because there are many possible ways of distributing the four
fundamental flavours from the D6-branes among the different factors in a
general gauge group. However, let us stress that once such a choice has
been made the condition $\b=0$ for all factors determines uniquely the
number of D4-branes stretched between the NS5-branes. There is also no
difficulty in writing down the spectrum.

Since our treatment in Section 5 will not be of direct application to
these models, and we will have little to say about their curves, we restrict
ourselves to just mentioning two interesting examples with simple gauge
group.

a) If only one NS5-brane is located, intersecting one of the orientifold
planes, and $2k$ D4-branes go around the circle in $x^6$, the gauge group
is $Sp(k)$ and there is a hypermultiplet in the antisymmetric. There are
also four fundamental flavours, coming from the presence of the D6-branes.
This theory has vanishing $\beta$ function, and
has received much attention \cite{aharony,dls,piljin} since it appears
as the world-volume theory of
multiple D3-brane probes \cite{bds} in the F-theory orientifold background
introduced by Sen \cite{senfth}.

b) A second example is to consider two unpaired NS5-branes, each at
one of the \mbox{$O6^-$'s}, and $n$ D4-branes (plus mirrors) suspended
between them. The resulting theory has gauge group $SU(n)$
and
there are two antisymmetric representations and four fundamental flavours.
Again the $\beta$ function vanishes. Notice that the same remarks as
in the previous subsection about Higgssing $SU(2k)$ or $SU(2k+1)$ to
$Sp(k)$ apply to this example.

\subsection{Possibility of introducing a global mass}

As mentioned above, a natural question that arises is whether it is
possible or not to introduce a global mass in these theories. This is
interesting because several $N=4$ theories belong to the class of models
one can analyze using these constructions. The global mass corresponds, in
these cases, to the adjoint hypermultiplet mass. For other theories, e.g.
those presented in subsection 3.2, it corresponds to a mass for the two
index antisymmetric matter, and in the general case it is simply the
sum of all the hypermultiplet masses (the masses of the fundamental
representations, when
they are present, must not be included in this counting, since their origin is
in the positions of
D6-branes, which are completely unconstrained), as in Section 2.
In any event, it constitutes an interesting deformation of the theories
under study.

The claim is that the mechanism introduced by Witten \cite{witten4d}, and
reviewed in Section 2, of twisting the M-theory spacetime so that the
D4-branes come back to a shifted value of $v$ while they go around $x^6$,
can be implemented in these configurations. In Section 5 we will show that
the ingredients required to introduce the twisting responsible for the
mass are
$Z_2$ invariant. From the point of view of the type IIA configurations we
have introduced, this is reflected in the fact that a net shift in the
positions of the D4-branes to the left and right of a NS5-brane is
compatible with the $Z_2$ symmetry, as shown pictorically in Figure~1,
for a particular example.

\setlength{\unitlength}{.7mm}
\begin{picture}(200,85)

\put(50,20){\line(0,1){60}}

\multiput(50,25)(0,17.5){3}{\qbezier(0,0)(22,0)(30,7.5)
\qbezier(30,7.5)(38,15)(15,15)
\qbezier(15,15)(12,15)(2.5,14)
\qbezier(-2.5,13.5)(-11,12)(-15,7.5)
\qbezier(-15,7.5)(-22,0)(0,0)}

\put(48.8,41.3){x}
\put(48.1,40.5){O}
\put(50.5,45){\scriptsize $O6^-$}

\put(64,56.3){x}
\put(63.8,55.9){\scriptsize{O}}
\put(70,59){\tiny $O6^+$}

\put(150,20){\line(0,1){60}}

\multiput(150,28)(0,17.5){3}{\qbezier(0,0)(22,0)(30,5)
\qbezier(30,5)(38,10)(15,10)
\qbezier(15,10)(12,10)(1.5,8)
\qbezier(-1.5,7.5)(-11,4)(-15,0)
\qbezier(-15,0)(-22,-7.5)(0,-7.5)
}
\put(152,21.5){\large{\}} {\small m}}

\put(148.9,40.3){x}
\put(148.2,39.5){O}
\put(140,42){\scriptsize $O6^-$}

\put(163,54.3){x}
\put(162.8,53.9){\scriptsize{O}}
\put(161,58){\tiny $O6^+$}

\put(100,30){\put(0,0){\line(0,1){10}}
\put(0,0){\qbezier(0,0)(6,0)(12,3)}
\put(2,8){\scriptsize{$v$}}
\put(7,-3){\scriptsize{$x^6$}}
\put(0,10){\vector(0,1){3}}
\put(10,2){\vector(2,1){3}}
}

\put(20,70){a)}
\put(120,70){b)}
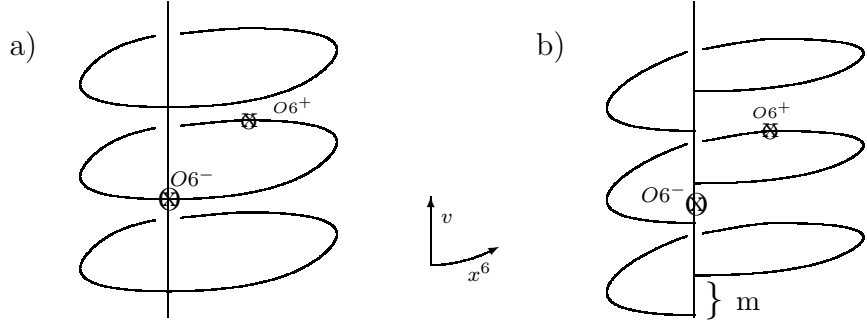
\put(15,7){Figure 1: Brane configurations in a particular
elliptic model}
\put(20,-3){with vanishing (a) and non-vanishing (b) global mass.}

\end{picture}

\bigskip
\bigskip

\section{On our way to M-theory}

In order to embed in M-theory the brane configurations discussed in the
last section, we introduce the new dimension $x^{10}$, which parametrizes a
circle. As in the models of Section 2, a torus appears naturally
in the M-theory spacetime, parametrized by $x^6$ and $x^{10}$.
The $(-1)^{F_L}$ operation in the $Z_2$ orientifold action is geometrized
as the inversion of $x^{10}$. The action of the $Z_2$ on spacetime is
$v\to -v$, $x^6\to -x^6$, $x^{10}\to - x^{10}$, so
one is quotienting by the natural $Z_2$
involution of the torus. This is compatible with any complex structure,
and thus the $\tau$ parameter survives in the orientifolded theory and
provides the microscopic gauge coupling of the field theory. When the
elliptic curve is written in the Weierstrass form (\ref{weierstrass}),
the involution corresponds to $y\to -y$ keeping $x$ invariant.

There are essentially two difficulties in finding the SW curve for these
theories in close analogy with the unitary case of Section 2. The first
concerns the ambient M-theory geometry in which the curve is embedded,
due to the presence of the O6-planes, and the D6-branes. As discussed in
the literature, these objects imply non-trivial geometries when lifted to
M-theory, namely, $S^1$ fibrations over the a base parametrized by
$x^6,v$.
A D6-brane is represented by a Taub-NUT space \cite{townsend, witten4d},
the $O6^-$ corresponds to an Atiyah-Hitchin space \cite{sw3d,senah,
landlop},
the $O6^+$ looks like a $D_4$ singularity in one of its complex structures
\cite{landlop,witfut}. Since our classical configurations include these
objects it is unclear that the M-theory geometries will reduce to such a
rigid structure as  $E_{\tau}\times \IC$, or a bundle thereof.

A second difficulty lies in the description of the curve the M5-brane is
wrapping. Looking at a prototypical gauge group in the families we are
studying, e.g. $Sp(k)$ $\times SU(2k+2)$ $ \ldots \times SU(2k+2p)$
$\times SO(2k+2p+2)$, we notice that different factors in the gauge group
have different numbers of D4-branes. This feature, as it stands, is
incompatible with a description of the SW curve as a spectral cover, which
would require a well defined number of sheets, independent of the $x^6$
position.
However, both difficulties are in fact related, and will be resolved at
once. In following subsections we discuss how the
geometry of M-theory can be simplified without changing the physical
properties described by the curve. This
will be shown in some simple examples with non-compact $x^6$. As a
by-product we will obtain a spectral cover interpretation for the curves
of theories with $\b=0$.

\subsection{Changing the geometry in simple examples}

The key observation is that sometimes a SW curve can be derived in
different ways which differ in the M-theory spacetime involved. This can
be illustrated in several examples with non-compact $x^6$, the simplest of
which is the Hanany-Witten effect \cite{hw}. The general fact that the
$x^6$ positions of D6-branes are not visible in the low energy gauge
theory allows the construction of e.g. the $N=2$ $SU(k)$ theory with $2k$
flavours in two ways, as depicted in Figure~2 for $k=2$.

\setlength{\unitlength}{.7mm}
\begin{center}
\begin{picture}(200,90)


\put(30,20){\line(0,1){60}}
\put(70,20){\line(0,1){60}}

\put(130,20){\line(0,1){60}}
\put(170,20){\line(0,1){60}}
\put(30,25){\line(1,0){40}}
\put(30,45){\line(1,0){40}}
\put(30,60){\line(1,0){40}}
\put(30,70){\line(1,0){40}}

\put(130,25){\line(1,0){40}}
\put(130,45){\line(1,0){40}}
\put(130,60){\line(1,0){40}}
\put(130,70){\line(1,0){40}}
\put(49,22){x}
\put(54,29){x}
\put(39,40){x}
\put(49,49){x}
\put(59,54){x}
\put(34,64){x}
\put(64,71){x}
\put(44,74){x}
\put(105,23){\line(1,0){25}}
\put(105,41){\line(1,0){25}}
\put(105,65){\line(1,0){25}}
\put(105,75){\line(1,0){25}}
\put(170,30){\line(1,0){25}}
\put(170,50){\line(1,0){25}}
\put(170,55){\line(1,0){25}}
\put(170,72){\line(1,0){25}}

\put(10,25){\line(0,1){10}}
\put(10,25){\line(1,0){10}}
\put(12,33){\scriptsize{$v$}}
\put(17,27){\scriptsize{$x^6$}}
\put(10,32){\vector(0,1){3}}
\put(17,25){\vector(1,0){3}}

\put(20,78){a)}
\put(95,78){b)}
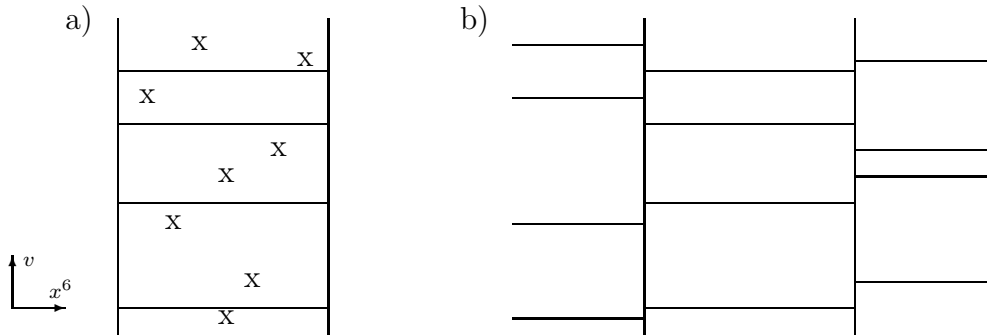
\put(20,10){Figure 2: Two realizations of $SU(4)$ with 8 flavours}
\end{picture}
\end{center}

In the first the flavours arise from $2k$ D6-branes. The M-theory spacetime
is a multicentered  Taub-NUT space, which can be written as
$t_1 t_2  = \prod_{i=1}^{2k} (v-m_i) $
in one of its complex structures. The SW curve is found to be
\cite{witten4d}
\beq
t_1^2 \; + \; p_k(v) t_1 \; + \; \prod_{i=1}^{2k} (v-m_i) \; = \; 0
\label{suk}
\eeq
In the second construction the D6-branes have been pushed to $\pm \infty$,
and flavours come from semi-infinite D4-branes. The
space parametrized by $x^6$, $x^{10}$ and $v$ is
`trivial' \footnote{Throughout this section we will refer to
this structure as `trivial' or `rigid', as opposed to more complicated
geometries.}, $\IR\times S^1\times
\IC$, which can be described as $t_1^{\prime} t_2^{\prime} = 1$. The curve
is
\beq
\prod_{i=1}^{k} (v-m_i)\; t_1^{\prime 2} \; + \; p_k(v)\; t_1^{\prime} \;
+ \;
\prod_{i=k+1}^{2k} (v-m_i) \; = \; 0
\label{sukprim}
\eeq

Both curves are equivalent through the change of variables
\beqa
t_1 \; = \; t_1^{\prime} \prod_{i=1}^k (v-m_i) \quad ; \quad
t_2 \; = \; t_2^{\prime} \prod_{i=k=1}^{2k} (v-m_i)
\label{cambio}
\eeqa

An unexpected payoff is that when one chooses the primed variables to
describe the curve, thus maximally simplifying the spacetime geometry, the
curve has a direct interpretation as a $k$-fold cover of the space
parametrized by $x^6$, $x^{10}$ if and only if the theory has $\b=0$
(see eq. (\ref{sukprim}) and Figure~2b).

\bigskip

We now turn to consider the case in which the RR charge which makes the
spacetime non-trivial comes from
orientifold six-planes. We are to show that even though in this case the
charged objects cannot be moved around, the change of variables leading to
a trivial spacetime and to well defined covers is still valid.

Consider
a brane configuration with an $O6^+$, and two
NS5-branes. The number of D4-branes we suspend between them is allowed to
be even or odd. Let us, for definiteness, discuss the even case, and place
$2k$ D4-branes joining the NS5-branes, arranged
in a $Z_2$ symmetric way. An example with $k=3$ is shown in Figure~3a.

\setlength{\unitlength}{.7mm}
\begin{center}
\begin{picture}(200,85)

\put(30,20){\line(0,1){60}}
\put(70,20){\line(0,1){60}}

\put(30,30){\line(1,0){40}}
\put(30,40){\line(1,0){40}}
\put(30,45){\line(1,0){40}}
\put(30,55){\line(1,0){40}}
\put(30,60){\line(1,0){40}}
\put(30,70){\line(1,0){40}}
\put(130,20){\line(0,1){60}}
\put(170,20){\line(0,1){60}}

\put(130,30){\line(1,0){40}}
\put(130,40){\line(1,0){40}}
\put(130,45){\line(1,0){40}}
\put(130,55){\line(1,0){40}}
\put(130,60){\line(1,0){40}}
\put(130,70){\line(1,0){40}}
\put(105,49.5){\line(1,0){25}}
\put(105,50.5){\line(1,0){25}}

\put(170,49.5){\line(1,0){25}}
\put(170,50.5){\line(1,0){25}}

\put(49,48.8){x}
\put(48.3,48){O}
\put(53,48.5){\scriptsize{$O6^+$}}
\put(149,48){\tiny{o}}
\put(152,48.5){\scriptsize{O6}}

\put(10,25){\line(0,1){10}}
\put(10,25){\line(1,0){10}}
\put(12,33){\scriptsize{$v$}}
\put(17,27){\scriptsize{$x^6$}}
\put(10,32){\vector(0,1){3}}
\put(17,25){\vector(1,0){3}}

\put(20,78){a)}
\put(100,78){b)}
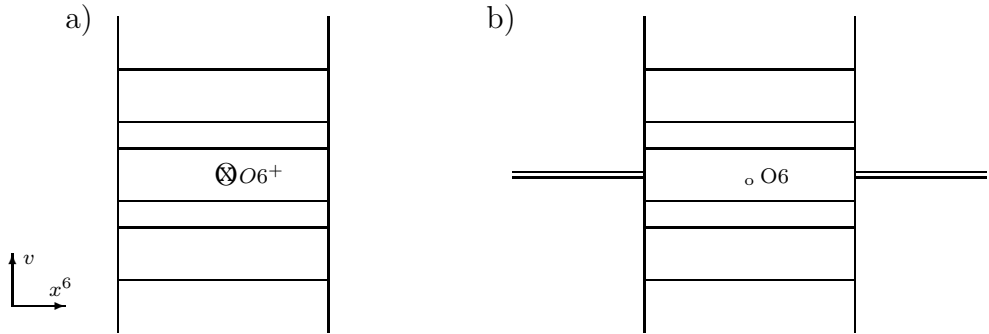
\put(-5,10){Figure 3: Pictorical representation of the change of variables
in the pure $SO(6)$ theory}

\end{picture}
\end{center}

The $N=2$ gauge theory
is the pure $SO(2k)$ (flavours are easily
added by introducing D6-branes. For notational convenience we do not
consider them, save for some comments at the end). The spacetime geometry,
as analyzed in \cite{landlop} is described by $t_1 t_2 = v^4 $
(with the $Z_2$ acting by interchanging $t_1$ and $t_2$ and inverting $v$)
and the curve is found to be
\beq
t_1^2 \; + \; p_k(v^2)\, t_1 \; + \; v^4 \; = \; 0
\label{socurve}
\eeq

Making the change of variables $t_1 = t_1^{\prime} v^2$,
$t_2=t_2^{\prime} v^2$,
the equation describing the ambient geometry is
$t_1^{\prime} t_2^{\prime}=1$ ($\IR\times S^1\times \IC$) and the curve is
\beq
v^2\; t_1^{\prime 2} \; + \; p_k(v^2)\; t_1^{\prime} \; + \; v^2 \; = \; 0
\eeq
which is pictorically represented in Figure~3b. It looks like if the
$+4$ charge of the orientifold had been pushed through the NS5-branes to
$\pm \infty$, creating new non-dynamical D4-branes in the process,
{\em \`a la} Hanany-Witten. The only remnant of the orientifold plane at
$v=0$ is the $Z_2$ quotient, all other effects, associated to its charge,
having been encoded in the presence of the extra semi-infinite D4-branes.

Notice that if we introduce $2k-2$ D6-branes (plus mirrors) in the
original configuration the additional flavours make the
$\b$-function vanish. The change of variables that makes spacetime
trivial gives a description of the curve in which the D6-branes have
been pushed through the NS5-branes, creating $2k-2$ D4-branes on the
outside the NS5-branes. The final curve is a $2k$-fold cover\footnote{We
are being slightly vague in our use of the word `cover'. E.g. even
(\ref{socurve}), which describes a theory with $\b\neq 0$, is a $2k$-fold
cover of the $t_1$-space, since, for each $t_1$ there are $2k$ roots for
$v$. However, as $t_1\to\pm\infty$ some of the roots diverge, and are not
interpreted as D4-branes but as the bending of the NS5-branes
\cite{witten4d}. To be precise, we will call covers those curves in which
the roots for $v$ remain finite even as $t_1\to\pm\infty.$} of $\IR\times
S^1$.

This exercise can also be performed for the $SO(2k+1)$ theory, yielding
analogous results.

\medskip

The same trick also works in the presence of an $O6^-$. Consider the brane
configuration in Figure~4a, leading to the pure $N=2$ Sp(k)
gauge theory.

\setlength{\unitlength}{.7mm}
\begin{center}
\begin{picture}(200,85)

\put(30,20){\line(0,1){60}}
\put(70,20){\line(0,1){60}}

\put(130,20){\line(0,1){60}}
\put(170,20){\line(0,1){60}}
\put(30,25){\line(1,0){40}}
\put(30,30){\line(1,0){40}}
\put(30,38){\line(1,0){40}}
\put(30,62){\line(1,0){40}}
\put(30,70){\line(1,0){40}}
\put(30,75){\line(1,0){40}}

\put(130,25){\line(1,0){40}}
\put(130,30){\line(1,0){40}}
\put(130,38){\line(1,0){40}}
\put(130,62){\line(1,0){40}}
\put(130,70){\line(1,0){40}}
\put(130,75){\line(1,0){40}}

\put(130,49){\line(1,0){40}}
\put(130,50.5){\line(1,0){40}}

\put(49,48.8){x}
\put(48.3,48){O}
\put(53,48.5){\scriptsize{$O6^-$}}

\put(149.5,49){\tiny{o}}
\put(150,52){\scriptsize{O6}}

\put(10,25){\line(0,1){10}}
\put(10,25){\line(1,0){10}}
\put(12,33){\scriptsize{$v$}}
\put(17,27){\scriptsize{$x^6$}}
\put(10,32){\vector(0,1){3}}
\put(17,25){\vector(1,0){3}}

\put(20,78){a)}
\put(100,78){b)}
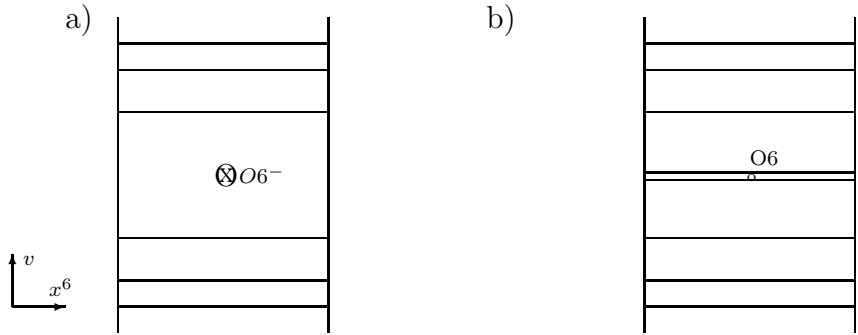
\put(-5,10){Figure 4: Pictorical representation of the change of variables
in the pure $Sp(3)$ theory}

\end{picture}
\end{center}

Following \cite{landlop} the curve can be obtained by taking
the spacetime to be $t_1 t_2 = v^{-4}$, which is `almost' the
Atiyah-Hitchin space, but for short distance corrections. The curve is
found to be
\beq
t_1^2 \; + \; [ p_k(v^2)\, +\, A v^{-2}]\; t_1 \; + \; v^{-4} \; = \; 0
\eeq
The  term proportional to $A$ takes into account the short distance
corrections not included before. The determination of the coefficient $A$,
which is not an order parameter, will be discussed in subsection
4.3 in a more general case.

Upon the change of variables $t_1 = t_1^{\prime} v^{-2}$, $t_2 = t_2^{\prime}
v^{-2}$, spacetime simplifies to $t_1^{\prime} t_2^{\prime}=1$, and the
curve becomes
\beq
t_1^{\prime 2} \; + [v^2 p_k(v^2)\, +\, A]\; t_1^{\prime} \; +\; 1 \; = \;
0
\label{spcurve}
\eeq
which is depicted in Figure~4b. In this case the D4-branes have
been created between the NS5-branes, as if charges from $\pm \infty$ had
come in to cancel the $-4$ charge of the orientifold.

Again, if one introduces $2k+2$
flavours to obtain $\b=0$, the change of variables simplifying the
spacetime transforms the curve in a $(2k+2)$-fold cover of $\IR\times
S^1$.

\medskip

Notice that the positions of the new D4-branes introduce a non-dynamical
parameter in the theories. In the case of the $O6^+$ this is fixed by the
condition that the D4-branes are at $v=0$. For the $O6^-$ case, it is
encoded in the coefficient $A$, which only depends on the dynamical scale
and masses in the theory. The new D4-branes are not simply frozen at $v=0$
in this last case. Behind this difference is the distinct behaviour of
both orientifold planes at strong coupling.

\subsection{More general examples}

In this subsection we generalize the results of the previous one for
theories with product gauge groups. Even though the direction $x^6$ is
still kept
non-compact, the picture obtained will be easy to implement in the
elliptic models. The treatment is mainly descriptive, even though there is
no difficulty in writing down the explicit formulae.

\medskip

Consider a brane configuration in the presence of an $O6^+$, yielding the
gauge group $SO(n)\times SU(n-2)\times \ldots \times SU(n-2l)$, and
hypermultiplets in bi-fundamentals, and with $n-2l-2$ fundamentals of the
last factor $SU(n-2l)$. All the factors have $\b=0$. This theory is
obtained by considering $l+1$ pairs of mirror NS5-branes, as well as the
appropriate number of D4-branes at each interval. The flavours in
the last factor are provided by $n-2l-2$ D6-branes
(and their mirrors). A typical example is depicted in Figure~5a ($l=1$,
$n=6$).

\setlength{\unitlength}{.7mm}
\begin{center}
\begin{picture}(200,85)

\put(20,20){\line(0,1){60}}
\put(40,20){\line(0,1){60}}
\put(60,20){\line(0,1){60}}
\put(80,20){\line(0,1){60}}

\put(20,37){\line(1,0){20}}
\put(20,42){\line(1,0){20}}
\put(20,45){\line(1,0){20}}
\put(20,55){\line(1,0){20}}

\put(40,25){\line(1,0){20}}
\put(40,35){\line(1,0){20}}
\put(40,40){\line(1,0){20}}
\put(40,60){\line(1,0){20}}
\put(40,65){\line(1,0){20}}
\put(40,75){\line(1,0){20}}

\put(60,45){\line(1,0){20}}
\put(60,55){\line(1,0){20}}
\put(60,58){\line(1,0){20}}
\put(60,63){\line(1,0){20}}

\put(24,29){x}
\put(34,62){x}
\put(64,36){x}
\put(74,69){x}
\put(49,48.8){x}
\put(48.3,48){O}
\put(45,54){\tiny{$O6^+$}}

\put(110,20){\line(0,1){60}}
\put(130,20){\line(0,1){60}}
\put(150,20){\line(0,1){60}}
\put(170,20){\line(0,1){60}}

\put(95,30){\line(1,0){15}}
\put(95,63){\line(1,0){15}}

\put(110,37){\line(1,0){20}}
\put(110,42){\line(1,0){20}}
\put(110,45){\line(1,0){20}}
\put(110,55){\line(1,0){20}}

\put(130,25){\line(1,0){20}}
\put(130,35){\line(1,0){20}}
\put(130,40){\line(1,0){20}}
\put(130,60){\line(1,0){20}}
\put(130,65){\line(1,0){20}}
\put(130,75){\line(1,0){20}}

\put(150,45){\line(1,0){20}}
\put(150,55){\line(1,0){20}}
\put(150,58){\line(1,0){20}}
\put(150,63){\line(1,0){20}}

\put(170,37){\line(1,0){15}}
\put(170,70){\line(1,0){15}}

\put(95,51.5){\line(1,0){15}}
\put(95,50.5){\line(1,0){35}}
\put(95,49.5){\line(1,0){35}}
\put(95,48.5){\line(1,0){15}}

\put(170,51.5){\line(1,0){15}}
\put(150,50.5){\line(1,0){35}}
\put(150,49.5){\line(1,0){35}}
\put(170,48.5){\line(1,0){15}}

\put(139,48.5){\tiny{o}}
\put(142,50.5){\tiny{O6}}

\put(10,78){a)}
\put(100,78){b)}
\put(5,10){Figure 5: Changing variables to get a cover in presence of an
$O6^+$}

\end{picture}
\end{center}

The curve for
this configuration can be easily written down following the recipe in
\cite{landlop}. Then one
performs the change of variables to eliminate the
RR charge coming from the orientifold and the D6-branes, so that the
ambient geometry
simplifies to a rigid $\IR\times S^1\times \IC_v$. Based on our experience
with simpler examples, we know that this corresponds to taking the $+4$
charge from the $O6^+$ and pushing it through the NS5-branes to $\pm
\infty$, and to performing the Hanany-Witten effect with the D6-branes. In
the process, new, non-dynamical D4-branes are created each time the charge
crosses a NS5-brane, and they appear precisely in the number required to
complete a total of $2k$ D4-branes in each interval between NS5-branes
(counting both the dynamical and non-dynamical branes), as made clear in
Figure~5b. The
final curve is a $2k$-fold cover of $\IR\times S^1$. Notice that the new
D4-branes created by the orientifold charge are frozen at $v=0$ (even
though in Figure~5b they are slightly separated, for clarity).

A similar argument works for the theories with gauge group $Sp(k)\times
SU(2k+2)\times \ldots \times SU(2k+2l)$ that arise in the presence of an
$O6^-$. The corresponding picture is shown in Figure~6. In this case the
D4-branes are created between the NS5-branes.

\setlength{\unitlength}{.7mm}
\begin{center}
\begin{picture}(200,85)

\put(20,20){\line(0,1){60}}
\put(40,20){\line(0,1){60}}
\put(60,20){\line(0,1){60}}
\put(80,20){\line(0,1){60}}

\put(20,30){\line(1,0){20}}
\put(20,35){\line(1,0){20}}
\put(20,42){\line(1,0){20}}
\put(20,70){\line(1,0){20}}

\put(40,25){\line(1,0){20}}
\put(40,75){\line(1,0){20}}

\put(60,30){\line(1,0){20}}
\put(60,58){\line(1,0){20}}
\put(60,65){\line(1,0){20}}
\put(60,70){\line(1,0){20}}

\put(29,22){x}
\put(24,24){x}
\put(29,44){x}
\put(34,54){x}
\put(22,66){x}
\put(29,59){x}

\put(69,39){x}
\put(76,32){x}
\put(64,44){x}
\put(69,54){x}
\put(74,74){x}
\put(69,76){x}

\put(49,48.8){x}
\put(48.3,48){O}
\put(45,54){\tiny{$O6^-$}}

\put(110,20){\line(0,1){60}}
\put(130,20){\line(0,1){60}}
\put(150,20){\line(0,1){60}}
\put(170,20){\line(0,1){60}}

\put(95,23){\line(1,0){15}}
\put(95,25){\line(1,0){15}}
\put(95,45){\line(1,0){15}}
\put(95,55){\line(1,0){15}}
\put(95,67){\line(1,0){15}}
\put(95,60){\line(1,0){15}}

\put(110,30){\line(1,0){20}}
\put(110,35){\line(1,0){20}}
\put(110,42){\line(1,0){20}}
\put(110,70){\line(1,0){20}}

\put(130,25){\line(1,0){20}}
\put(130,75){\line(1,0){20}}

\put(150,30){\line(1,0){20}}
\put(150,58){\line(1,0){20}}
\put(150,65){\line(1,0){20}}
\put(150,70){\line(1,0){20}}

\put(170,33){\line(1,0){15}}
\put(170,40){\line(1,0){15}}
\put(170,45){\line(1,0){15}}
\put(170,55){\line(1,0){15}}
\put(170,75){\line(1,0){15}}
\put(170,77){\line(1,0){15}}

\put(130,51.5){\line(1,0){20}}
\put(110,50.5){\line(1,0){60}}
\put(110,49.5){\line(1,0){60}}
\put(130,48.5){\line(1,0){20}}

\put(139,48.5){\scriptsize{O}}
\put(139,53){\tiny{O6}}

\put(10,78){a)}
\put(100,78){b)}
\put(5,10){Figure 6: Changing variables to get a cover in presence of an
$O6^-$.}

\label{figura5}
\end{picture}
\end{center}

The main conclusion of this section is the existence of certain variables
such that, when the curve is written in terms of them, it takes the form
of a cover of $\IR\times S^1$ with a well defined number of sheets. The
appearance of new D4-like tubes connecting the NS5-brane can be understood
from a Hanany-Witten effect associated to the charge of the orientifold
six-planes.

A similar analysis can also be performed in case there is an unpaired
NS5-brane intersecting the orientifold plane. The conclusion we have just
stated is unchanged.

In the following subsection presents a discussion of the precise
conditions that fix the new parameters
which are associated to the non-dynamical D4-branes created by the
orientifold charges.

\subsection{Determining the non-dynamical parameters}

As mentioned above, when dealing with brane configurations in the
presence of an \mbox{$O6^-$}, the non-dynamical D4-branes that appear
due to the
O6-plane charge are not simply frozen at $v=0$, their positions are
encoded in the coefficients of the terms which describe the strong
coupling splitting of the $O6^-$. A systematic way of determining these
coefficients was proposed in \cite{landlop}, by imposing that the equation
should define a curve in the Atiyah-Hitchin space. However, in their
original formulation these conditions are of little use when $x^6$ is
made compact. Our purpose in the following is to find the geometric
interpretation of these conditions, so that they can be implemented in
the construction of elliptic models. This can be done in a fairly general
way, with independence of the particular brane configuration one
considers.

Let us place an $O6^-$ at $v=0$, described by $t_1 t_2=v^{-4}$, and write
down the curve for a certain brane configuration, which involves a total
of, say, $n$ NS5-branes (counting also mirrors). Perform now the change of
variables $t_1=t_1^{\prime} v^{-2}$, $t_2=t_2^{\prime} v^{-2}$, so that
spacetime becomes trivial. Our
starting point is the curve thus obtained, which we may denote as
$F(t_1^{\prime},v)=0$. It is a polynomial, and has degree $n$ in
$t_1^{\prime}$. Notice that it must be
invariant under $t_1^{\prime} \to 1/t_1^{\prime}$, $v\to -v$, and that it
is a cover of $\IR\times S^1$ when the $\b$ function of all factors
vanishes.

In \cite{landlop} a few changes of variables are made in this expression.
First, make the change $t_1^{\prime} = vs_1^{\prime}$, then divide the
equation by $v^n$, yielding
$v^{-n}F(vs_1^{\prime},v)=0$. Now perform the change $s_1^{\prime}=
z_1^{\prime} -v^{-1}$, so one gets $v^{-n} F(vz_1^{\prime}-1,v)=0$.
Through these changes, the new coordinates can describe the Atiyah-Hitchin
space, and the condition that the equation defines a curve in this space
is that it should not contain negative powers of
$v$.  Equivalently, the function $G(z_1^{\prime},v)=F(vz_1^{\prime}-1,v)$
starts its Taylor expansion in $v$ with the $v^n$ term;
\beqa
\frac{\partial^l}{\partial v^l} G(z_1^{\prime},v) \big|_{v=0} \; = \; 0
\quad \quad ; \quad l=0, \ldots, n-1
\eeqa
By application of the chain rule, these conditions can be translated to
conditions on $F(t_1^{\prime},v)$:
\beqa
\frac{\partial^k}{\partial t_1^{\prime k-l} \partial v^{l}}
F(t_1^{\prime},v)
\big|_{\stackrel{t_1^{\prime}=-1}{v=0}} \; \; = \; 0 \quad \quad ; \quad
k=0,\ldots, n-1\; ;\; l=0,\ldots,k
\eeqa

Defining the functions $H_l(t_1^{\prime})$ on $\IR\times S^1$
\beqa
H_l(t_1^{\prime}) \; = \; \frac{\partial^l}{\partial v^l}
F(t_1^{\prime},v) \big|_{v=0} \quad ; \quad l=0,\ldots,n-1,
\label{defh}
\eeqa
the conditions state that $H_l(t_1^{\prime})$ has a zero of order $n-l$ at
the fixed point $t_1^{\prime}=-1$.

Geometrically, this simply means that the curve passes through that point $n$
times. This is made clear if we introduce a local variable
$t=t_1^{\prime}$ and write the local description of the cover near $v=0$,
$t=0$ as the factorized expression
\beqa
(v-r_1 t) (v-r_2 t) \cdots (v - r_n t) \; = \; 0.
\eeqa
The coefficients $r_i$ are generic, so that the intersection of the
different sheets is generically transverse.

\medskip

As a very simple example, we can consider equation (\ref{spcurve}), which
corresponds to a particular case of $F(t_1^{\prime},v)=0$ (simply note its
symmetry). The condition to fix $A$ is that $F(t_1^{\prime},0)$ and
$\frac{\partial}{\partial v} \, F(t_1^{\prime},v)|_{v=0}$ should have a
double
and a simple zero, respectively, at $t_1^{\prime}=-1$. This fixes the
value $A=-2$. The coefficients in \cite{landlop} are very efficiently
reproduced by using these conditions, even in more complicated examples.
They also will turn out to be crucial in the definition of the
covers in the elliptic models of Section 5.

\bigskip

{\large {\em The integrable system}} \footnote{I am thankful to E.~Witten
for explaining the contents of this subsection to me.}

To end this subsection, we briefly comment on an essential property of the
conditions we have found, namely the family of curves obtained
has the structure of an integrable system. This is a necessary
condition if they are to describe the Coulomb branch of a $N=2$ gauge
theory \cite{dw, gorsky,marwar, ponjas} (see \cite{integr} for
a recent overview of this connection, and for a more complete list
of references). As discussed in \cite{witten4d}, many complex integrable
systems can be constructed as the deformation space of a certain curve
(and with a line bundle over it) embedded on a two dimensional complex
symplectic manifold \cite{del}. In our case, the curve is the
Riemann surface the M5-brane is wrapping, and the bundle is specified by a
point in the jacobian of the curve. The embeding space is the quotient of
$\IR\times S^1\times \IC_v$ by the $Z_2$ symmetry.

The integrable system structure appears only if all deformations
of the curve (consistent with the asymptotic behaviour) are included in
the family obtained by varying the order parameters. Since the curves in
our family
are forced to pass through a given point, it seems at first sight that
there are deformations we have not included, and that the family of curves
we have constructed does not lead to an integrable system. However, this
point is an $A_1$ singularity, since it is a fixed point under the $Z_2$
action, and must be
blown up. In the resolved space, the conditions amount to
especifying the intersection number of the (proper transform of the)
curve with the exceptional divisor. This number depends only on the
homology class of the curve, and is
stable under deformations. The family of curves describes a whole
deformation class with the resolution of $(\IR\times S^1\times \IC)/Z_2$
as ambient space, and leads to an integrable system.

The construction of the curves for the orientifolded models can be
understood as a restriction in the parameters in the curves for gauge
theories with unitary gauge groups. Especifically, the restrictions
correspond to the $Z_2$ symmetry and to forcing the curve to pass through
a given point. Let us remark that it is possible to construct the
integrable system for the orientifolded theory as an appropriate
restriction of the larger integrable system for the theory with unitary
groups \cite{don1}

In Section 5 we will construct curves in the context of elliptic models
which verify the same kind of conditions, and this argument showing
their integrable system structure also applies.

\subsection{The NS5-brane intersecting the $O6^-$}

There is still an important lesson to learn from these non-compact
examples. We have just seen that the presence of an $O6^-$ is encoded in
the curve by the condition that it must pass through a fixed point at a
certain order. This point corresponds in the classical limit to the
location of the $O6^-$. However, there are two fixed points on the
$S^1$ that shrinks to this classical location. In a general situation
there is no difference in forcing the curve to pass through one or the
other.

However, there is another object that can appear at the classical value of
$x^6$ corresponding to the $O6^-$, a NS5-brane. In the M-theory picture,
its
position corresponds to a fixed point in $\IR\times S^1$ at which the
curve goes to infinity in $v$. In this situation, it makes sense to ask
whether the fixed point associated to the NS5-brane coincides or not with
the fixed point of the $O6^-$ (the one through which the curve passes).

Let us answer this question by discussing a concrete example. Consider the
gauge theory $SU(N)$ with matter in the antisymmetric representation. It
is realized by a symmetric distribution of three NS5-branes in presence of
an $O6^-$ (one NS5-brane intersects it). We also let $N$ D4-branes be
suspended at each interval in $x^6$, again respecting the symmetry. The
curve describing the theory is \cite{landlop}
\beqa
t_1^{\prime \, 3} \; + \; t_1^{\prime\, 2}\, \big[ v^2 \,\prod_{i=1}^N
(v-a_i) \, + \, 3 \big] \;
+ \; t_1^{\prime} \, \big[ (-1)^N\, v^2\,\prod_{i=1}^N (v+a_i) \, + 3 \big]
\; + \; 1 \; = \; 0
\label{triple}
\eeqa
As mentioned above, the zeroes have been imposed at $t_1^{\prime}=-1$. To
detect the location of the NS5-brane, we isolate the dominant terms as
$v\to\infty$, and find the solution $t_1^{\prime}=-(-1)^N$. For even
values of $N$
this coincides with the previous fixed point, while for odd $N$ it is
different.

This fact is of utmost importance in the context of elliptic models,
since it will allow for a nice geometric interpretation of Montonen-Olive
duality \cite{mo} in terms of the SW curves. It may also find other
applications in explaining field theory phenomena for which the parity in
the number of colors makes a crucial difference.

\section{Solution of the $O6^+$-$O6^-$ elliptic models with vanishing
global mass}

The purpose of this section is to describe the SW curves in the
$O6^+$-$O6^-$ orientifold background
for the case of vanishing global mass. The theories in which
$x^6$ is compact can be obtained by putting together two theories with
non-compact
$x^6$, so many results we learnt in Section~4 can be applied to
analyze the elliptic models.
We turn to proposing the  solutions for the four families of models
introduced in section 3.1.

\subsection{Models in family i)}

Let us start by studying the first class of models. A general
theory in this family has gauge group $Sp(k)$$\times
SU(2k+2)$$\times\ldots$$\times SU(2k+2p)$$\times SO(2k+2p+2)$, and
hypermultiplets in bi-fundamental representations. The ambient spacetime
in which the curve is embedded can be taken to be $E_{\tau}\times \IC_v$,
if the RR charge of the $O6^+$ is pushed through the NS5-branes
until it cancels against the charge of the $O6^-$. In the process, new
D4-branes, which are non-dynamical, are created, precisely in the number
appropriate to make the final curve a
$(2k+2p+2)$-fold cover of $E_{\tau}$. This is described by the equations
\beqa
y^2  & = & (x-e_1(\tau)) (x-e_2(\tau))(x-e_3(\tau)) \\
v^{2k+2p+2} & +  & f_1(x,y)  v^{2k+2p+1} + \ldots + f_{2k+2p+1} (x,y)\; v
+ f_{2k+2p+2} (x,y) = 0 \nonumber
\label{coverorient}
\eeqa
The first equation describes the M-theory spacetime, and the second,
the SW
curve. This last plays the role of the equation $F(t_1^{\prime},v)=0$ in
subsection 4.3, a fact we are to use soon.

We have to specify the conditions on the elliptic functions $f_i(x,y)$.

{\bf a)} First, the equations must be invariant under the $Z_2$ action,
$y\to -y$, $v\to -v$. This implies that the functions $f_{2l}(x,y)$ must
be even under $y\to -y$, and the $f_{2l+1}(x,y)$, odd.

\medskip

Let us pause for a moment to consider the implications of this
projection
on the possibility of turning on a global mass. As shown in reference
\cite{witten4d}, and stressed in Section 2, the global mass appears as a
simple pole of $f_1(x,y)$ at infinity, typically represented as a behaviour of
the form $m y/x$,
which is odd in $y$. This is precisely the parity selected by the
orientifold projection, and so, as anticipated pictorically in Section 3,
the global mass is compatible with the $Z_2$ projection.

\medskip

{\bf b)} The functions $f_i(x,y)$ have simple poles at the locations of
the NS5-branes, and are holomorphic at infinity. Recalling a few facts
about certain simple elliptic functions we can be rather explicit in
describing the functions $f_i(x,y)$, and count how many parameters they
depend on. So let us note that
the monomial $x-a$, for $a\neq e_i$, has two simple zeroes at mirror
locations
($x=a$, and $y$ given by the two solutions of the Weierstrass equation)
\footnote{If $a$ is a fixed point, $x-a$ has a double zero at the
corresponding point.}, and a double pole at infinity. The function $y$ has
simple zeroes at the three points $y=0$, $x=e_1,e_2,e_3$, and a
triple pole at infinity.

If we denote the locations of the $p+1$ NS5-branes (plus mirrors) by
$x=x_1,\ldots, x_{p+1}$, the structure of the functions $f_i(x,y)$ is
\beqa
f_{2l}(x,y)\; & = &\; C_l\; \frac{(x-a_1^l)\ldots (x-a_{p+1}^l)}
{(x-x_1)\ldots (x-x_{p+1})} \nonumber\\
f_{2l+1}(x,y)\; & = &\; D_l\; y\; \frac{(x-b_1^l)\ldots (x-b_{p-1}^l)}
{(x-x_1)\ldots (x-x_{p+1})}
\label{exp1}
\eeqa
The even functions depend on $p+2$ parameters (the $p+1$ zeroes and an
overall constant), and the odd ones, on $p$ (the $p-1$ zeroes and a
normalization).

These expressions are particular versions of eq.(\ref{sigmas}) for which
the parity properties facilitate expressing the Weierstrass sigma
functions in terms of $x$ and $y$.

\medskip

{\bf c)} The last condition comes from the analysis in subsection 4.3,
and concerns the freezing of the extra parameters associated
to the non-dynamical D4-branes. It comes out
that the conditions as stated after eq.(\ref{defh}) are easily implemented.
Noticing that the second equation in (\ref{coverorient}) is the
analogous of $F(t_1^{\prime},v)=0$, the function $H_l(t_1^{\prime})$ in
(\ref{defh}), obtained by
taking the $l$-th derivative with respect to $v$ at $v=0$ is
precisely the analogous of $f_{2k+2p+2-l}$. Since the number of
NS5-branes is $2p+2$, the conditions amount to imposing $f_{2k+l}(x,y)$
to have a zero of order $l$ at a fixed point. Equivalently, these
conditions state that the curve passes $2p+2$ times through this point.
The point defines the position of the $O6^-$ in the classical limit. In
the following we will denote its location in $E_{\tau}$ by $P_{O6^-}$

From the expressions (\ref{exp1}) we see that the even functions
$f_{2k+2p+2}$, $f_{2k+2p}$, $\ldots, f_{2k+2}$ adquire zeroes of the
desired multiplicity upon tuning $\sum_{l=1}^{p+1}l=(p+1)(p+2)/2$ of
their parameters. The
zeroes in the odd functions $f_{2k+2p+1}$, $f_{2k+2p-1}$, $\ldots,
f_{2k+1}$ are obtained upon fixing $p(p+1)/2$ parameters (recall that
$y$ already provides simple zeroes at the fixed points).

\bigskip

This analysis allows us to compute the number of parameters in the
spectral cover, $(k+p+1)(p+2)$$+(k+p+1)p$ are initially present in
(\ref{exp1}) and $(p+1)(p+2)/2$$+p(p+1)/2$ are removed by the
additional conditions.
The total amounts to
$2kp+p^2+2p+2k+1$ parameters. This is to be compared with
the number of parameters in the gauge theory. We have $p$ mass parameters,
as well as $k$$+\sum_{l=1}^p (2k+2l-1)$$+(k+p+1)$ order parameters, adding
to a total of $2kp+p^2+2p+2k+1$, and matching the previous result.

\medskip

As a further check on these curves, let us consider the theory
with gauge group $Sp(k)\times SO(2k+2)$ and matter in the bi-fundamental.
The curve for this model can be found as a particular example of the
elliptic models in \cite{lll}, in which an orientifold {\em four}-plane
was included. We can reproduce this curve from our construction for $p=0$
as follows. We must consider a $(2k+2)$-fold cover of $E_{\tau}$, in
which at most simple poles are allowed at the locations of the
NS5-branes. One easily checks the odd elliptic functions cannot satisfy
that and are forced to
vanish, so that only even powers of $v$ appear in the cover, which then
has the structure
\beqa
v^{2k+2} + f_2(x,y) v^{2k} +\ldots + f_{2k-2}(x,y) v^2 + f_{2k}(x,y) = 0
\label{elll}
\eeqa
Moreover, one
must impose (as discussed before) that $f_{2k+2}(x,y)$ must have a double
zero at $P_{O6^-}$. The corresponding curve matches precisely
that proposed in
\cite{lll}, even though the conditions on the elliptic functions come from
different physical mechanisms \footnote{Actually, some of the properties
that characterize our elliptic functions are different from those in
\cite{lll} (e.g. their functions are not {\em a priori} constrained to be
even). However, upon going to the explicit expressions for these
functions they come out to be identical in both constructions (e.g. the
only functions that satisfy the requirements in \cite{lll} are, {\em a
posteriori}, even)}.

In the remainder of this
section we will discuss the structure of the curves for the ii)
iii) and iv) families, stressing the novel features that arise, and
reproducing the flat directions which connect different models.

\subsection{Models in family ii)}

The general gauge group for the second family of models is $SU(n)$$\times
SU(n+2)$$\times\ldots$$\times$ $SU(n+2p)$$\times SO(n+2p+2)$. This
suggests to take a $(n+2p+2)$-fold cover of $E_{\tau}$ as the SW curve.
The
$Z_2$ invariance implies that the functions $f_{2l}(x,y)$ must be even
under $y\to -y$, and the functions $f_{2l+1}(x,y)$, odd. Notice that this
is so
regardless whether $n$ is even or odd.

Besides the $p+1$ pairs of simple poles, they must
have a simple pole at a fixed point, say $x=e_1$, since there is a
NS5-brane
there. This yields the following structure for the elliptic functions
\beqa
f_{2l}(x,y)\; & = & \; C_l \;\frac{(x-a_1^l)\ldots (x-a_{p+1}^l)}
{(x-x_1)\ldots (x-x_{p+1})} \nonumber\\
f_{2l+1}(x,y)\; & = & \; D_l\; y \;\frac{(x-b_1^l)\ldots (x-b_{p}^l)}
{(x-x_1)\ldots (x-x_{p+1})(x-e_1)}
\label{exp2}
\eeqa
The even functions depend on $p+2$
parameters, and have the peculiarity that they actually are holomorphic at
the location of $O6^-$. The odd functions depend on $p+1$
parameters and have a pole there.

In order to discuss the conditions on the multiplicity of zeroes at
$P_{O6^-}$, we should distinguish between the even and odd $n$
cases. Recall, from our findings in subsection 4.4, that they differ in
the relative positions of $P_{O6^-}$ and $x=e_1$, the location of the
NS5-brane pole.

\medskip

{\large {\em Even n}}

For $n=2k$, the point $P_{O6^-}$ must coincide with $x=e_1$. The SW curve
is a
$(2k+2p+2)$-fold cover of $E_{\tau}$, as in eq. (5.1). Since there are
$2p+3$ NS5-branes, the function
$f_{2k+l}(x,y)$ must have a zero of order $l+1$ at $x=e_1$. However, since
it is in principle allowed to have a pole there,
only a {\em net} zero of order $l$ should be imposed. For the even
functions $f_{2k+2p+2}$, $f_{2k+2p}$, $\ldots f_{2k}$ the desired
multiplicities require fixing $(p+1)(p+2)/2$ parameters. The odd functions
$f_{2k+2p+1}$, $f_{2k+2p-1}$, $\ldots f_{2k+1}$
require fixing $(p+1)(p+2)/2$ parameters as well.

In total, the spectral cover depends on $2kp+p^2+2p+3k+1$ parameters, in
agreement with the result from the counting in the gauge theory (there
are $2kp+p^2+p+3k$ Coulomb moduli and $p+1$ masses).

As a particular but rather trivial example, let us consider the $p=-1$
case, which as explained in Section~3 corresponds to the $N=4$ $SO(2k)$
theory (with vanishing adjoint mass for the moment). The
functions $f_l(x,y)$ in the equation for the $2k$-cover of $E_{\tau}$ have
at most one simple pole, but an elliptic function
cannot have just one simple pole, so they must be holomorphic in
$E_{\tau}$ and thus constant. For the odd functions this
constant is zero, so the equation for the cover reduces to a polynomial in
$v^2$ with constant coefficients, that can be factorized as
\beq
\prod_{i=1}^k  (v^2 - v_i^2) \; = \; 0.
\eeq
This corresponds to $k$ pairs of copies of $E_{\tau}$, each giving the
$\tau$ parameter of a $U(1)$ on the Coulomb branch, the correct
SW curve for the $N=4$ $SO(2k)$ theory.

\medskip

Before moving on to the odd $n$ theories, let us illustrate how the
Higgssing of $SU(2k)$ with the antisymmetric representation to $Sp(k)$
proceeds in the SW curve. It corresponds to the removal of the NS5-brane
intersecting the $O6^-$, so, in analogy with a similar process studied in
Section 2, we let one of the zeroes of each elliptic function match the
pole associated to the NS5-brane at $x=e_1$. However, we have already
noticed that
the even functions do not have such a pole, and neither do some of the odd
ones, due to the conditions of having zeroes there. The Higgssing only
requires forcing new simple zeroes in
$f_1, f_3,\ldots, f_{2k-1}$, i.e. we must tune $k$ parameters. This is
precisely
the number of parameters lost in the Higgssing in the gauge theory, $k-1$
coming from Coulomb branch moduli, and the last being the
antisymmetric hypermultiplet mass. It is also straightforward to check
that the spectral cover of the family ii) becomes a curve associated to a
model of type i).

\medskip

{\large {\em Odd n}}

When $n$ is odd, say $2k+1$, the spectral cover has $2k+2p+3$ sheets, and
is given by an equation
\beqa
v^{2k+2p+3} + f_1(x,y)\; v^{2k+2p+2} + \ldots + f_{2k+2p+2}(x,y)\; v +
f_{2k+2p+3}(x,y) \; =\; 0
\label{coverorient2}
\eeqa
In this case $P_{O6^-}$ differs from $x=e_1$. We set it e.g. at $x=e_2$.
The classical limit in which the NS5-brane and the $O6^-$ share the same
location is recovered by letting $x=e_1$ and $x=e_2$ coalesce. Other
classical limits, related to this by $SL(2,\IZ)$ can be achieved, as will
be described in subsection 5.5.

The condition of the curve passing through $x=e_2$ at order $2p+3$ implies
that $f_{2k+l}(x,y)$ has a zero of order $l$ at this point. This can be
achieved by fixing a total of $(p+1)(p+2)$ parameters.

Note the interesting property that, since we have imposed vanishing
conditions on $f_{2k+2p+3}$ in enough quantity to eliminate all its
parameters, it identically vanishes. Thus a whole sheet in the cover
is stuck at $v=0$, and a $v$ factors out in eq.(\ref{coverorient2}). In
fact
such phenomenon is already present in similar cases with non-compact
$x^6$.

As for the counting of free parameters, the cover finally contains
$2kp+p^2+3p+3k+2$, in agreement with the number in the gauge field theory,
as one can easily check.

By a reasoning analogous to that of the even $n$ case, the curve which is
obtained for the $N=4$ $SO(2k+1)$ theory consists of a polynomial that can
be factorized as
\beq
v\, \prod_{i=1}^k  (v^2 - v_i^2) \; = \; 0.
\eeq
It provides the correct description for the massless adjoint case, $2k+1$
copies of $E_{\tau}$ distributed in a $Z_2$ invariant fashion.

Let us finally recall that these models have a complicated flat direction
that connects them to theories of type i), through the breaking
(\ref{comphig1}). It is associated to the removal of the NS5-brane at
$x=e_1$, and a D4-brane. This is reproduced in the curves by forcing an
additional zero at $x=e_1$ in all the odd functions, except for
$f_{2k+2p+3}$, which is identically zero and has no parameters left.

The whole process requires tuning $k+p+1$ parameters, agreeing with the
gauge theory counting: the rank of the group is lowered by $k+p$ units and
one mass parameter is lost. On the final equation for the cover one should
forget the $v$ factor (removal of the D4-brane) in order to obtain a curve
in the family i). Notice that in the matching of the number of parameters
it is crucial that
the pole to be eliminated is not at the same position as $P_{O6^-}$. This
provides a nice check of the phenomenon we uncovered in the non-compact
$x^6$ examples in section 4.4 and translated to the elliptic models.

\subsection{Models in family iii)}

The models in this family have gauge group $Sp(k)$$\times
SU(2k+2)$$\times\ldots$$\times SU(2k+2p)$$\times SU(2k+2p+2)$. The SW
curve is a $(2k+2p+2)$-fold cover of $E_{\tau}$, with the elliptic
function
having the same parity restrictions as in previous cases. They must have
simple poles at the $p+1$ mirror positions of the NS5-brane, and at a
fixed point. If this last is chosen at $x=e_1$, the elliptic
functions are given by the expressions (\ref{exp2}). So again each even
function depends on $p+2$ parameters, and each odd one, on $p+1$.

Since the classical position of the NS5-brane coincides with that of the
$O6^+$, it follows that it differs from that of $P_{O6^-}$. We choose this
last to be e.g. $x=e_2$. The classical limit is recovered by letting $e_2$
and $e_3$ coalesce. The conditions on the multiplicity of zeroes at
$P_{O6^-}$
are that $f_{2k+l}$ must have a zero of order $l$ at that point.
For the even functions this
requires fixing $(p+1)(p+2)/2$ parameters, and for the odd ones,
$p(p+1)/2$. The spectral cover comes out to
depend on $2kp+p^2+3k+3p+2$ parameters, in agreement with the counting in
the gauge theory.

The case $p=-1$ corresponds to the $N=4$ $Sp(k)$ theory, again with zero
adjoint mass. By a reasoning analogous to that above, the SW curve our
configurations produce consists of a set of $2k$ copies of $E_{\tau}$ at
symmetrical values in $v$.

In this case the Higgssing to a model of type i) is accomplished by
forcing a zero at $x=e_1$ in all the odd functions. This
requires the freezing of $k+p+1$ parameters (matching the gauge theory
result) and yielding a spectral cover of type i).

Finally, let us mention that the covers described in this subsection are
identical to those in the family ii) with and odd number of sheets (they
only differ in a D4-brane stuck at $v=0$). This will be interpreted as a
manifestation of Montonen-Olived duality, as argued in section 5.5.

\subsection{Models in family iv)}

The final case, with gauge group $SU(n)$$\times
SU(n+2)$$\times\ldots$$\times SU(n+2p)$$\times SU(n+2p+2)$, is a
mixture of cases ii) and iii). The SW curve is a $(n+2p+2)$-fold cover
with the elliptic functions having the structure
\beqa
f_{2l}(x,y)\; & = & \; C_l\; \frac{(x-a_1^l)\ldots (x-a_{p+1}^l)}
{(x-x_1)\ldots (x-x_{p+1})} \nonumber\\
f_{2l+1}(x,y)\; & = & \; D_l\; y\; \frac{(x-b_1^l)\ldots (x-b_{p+1}^l)}
{(x-x_1)\ldots (x-x_{p+1})(x-e_1)(x-e_3)}
\label{exp3}
\eeqa
if the poles of the NS5-branes are located at $x=e_1,e_2$. We see that
each function depends on $p+2$ parameters.

Again, to discuss the orders of vanishing at the $O6^-$ we distinguish the
even and odd $n$ cases.

\smallskip

{\large {\em Even n}}

For $n=2k$ the point $P_{O6^-}$ must coincide with one of the poles, say
$e_1$. The
conditions are that the function $f_{2k+l}(x,y)$ must have a {\em net}
zero of order $l$ at such point. This fixes a total of $(p+1)(p+2)$
parameters, and the spectral
cover depends on $2kp+p^2+3p+4k+2$ parameters, again in agreement with the
computation in the gauge theory.

There is a possible Higgsing to models of type ii) through the breaking of
$SU(2k+2p+2)$
to $SO(2k+2p+2)$ by a vev for the symmetric hypermultiplet. It is
accomplished by forcing a zero at $x=e_3$ in all the odd functions, in
order to eliminate the pole. In the
process, $k+p+1$ parameters are lost. This counting matches the
result from the gauge theory.

There is also a Higgssing to models of type iii), breaking the $SU(2k)$ to
$Sp(k)$ by a antisymmetric hypermultiplet vev. It is reproduced in the
equations by eliminating the pole at $x=e_1$ imposing the presence of a
zero there in $f_1$,$f_3$,$\ldots f_{2k-1}$. In the process one fixes $k$
parameters, as should be the case.

\smallskip

{\large{\em Odd n}}

When $n$ is odd, say $2k+1$, the spectral cover takes the form in
eq.(\ref{coverorient2}). In this case $P_{O6^-}$ must not coincide with
the poles, and we set it e.g. at $x=e_2$. The function $f_{2k+l}(x,y)$
must have
a zero of order $l$ at that point. This requires fixing a total of
$(p+1)(p+2)$ parameters. Notice that $f_{2k+2p+3}$ does not identically
vanish, since it still conserves one degree of freedom.

In total, the curve depends on $2kp+p^2+4k+4p+4$ parameters, matching the
result from the gauge theory.

In order to reproduce the flat direction breaking to theories of the
family ii), one just forces zeroes to eliminate one of the poles in all
the odd functions. Since both poles
are essentially equivalent in the M-theory description, they only
differ in their relation with $P_{O6^-}$ in a particular classical limit.
If we consider the weak coupling limit in which $e_1$ coalesces with $e_2$,
the pole to be removed to recover a type ii) model is that at $x=e_3$.
This can be done by fixing $k+p+2$ parameters, as is
required to agree with the gauge theory breaking of $SU(2k+2p+3)$ to
$SO(2k+2p+3)$ using the symmetric hypermultiplet. Notice that also the
resulting spectral cover fits nicely in the class describing type ii)
models. In particular, the last parameter in $f_{2k+2p+3}$ has been
frozen, so this function ends up as identically vanishing, as must be the
case for theories in the family ii).

The flat direction breaking to theories in the family iii), with Higgssing
given by eq.(\ref{comphig2}), is reproduced by forcing an additional zero
at $x=e_1$ in all the odd functions. This requires tuning $k+p+2$
parameters, which is the correct gauge theory result. Note that in this
case $f_{2k+2p+3}$ also vanishes identically at the end, and the
equation for the cover has a $v$ factor. This is associated to the
D4-brane to be removed, and should be forgotten. The remaining equation
describes the SW curve for  a type iii) theory.

If the classical limit is defined in a different way, e.g. as the
coalescence of $e_3$ and $e_2$, the weak coupling interpretation of the
flat directions we have described is interchanged. This is again a
manifestation
of the Montonen-Olive duality in our curves, as we comment in some more
detail in section 5.5.

\bigskip

As a particular example for which the curve can be compared with a known
answer, let us consider the case $p=-1$, $n=2$. The gauge theory is
$SU(2)$, and it has matter transforming as an adjoint and a singlet. They
have opposite masses, so that their sum vanishes. Since the singlet is
decoupled, our curve should reproduce the answer found in \cite{swsu2} fo
the $SU(2)$ theory with an adjoint or arbitrary mass.

The equation for our cover is given by
\beqa
v^2 \; + \; \frac{y\,m\,(e_1-e_2)}{(x-e_1)(x-e_2)}\, v\;+\; u\; =\;0
\label{su2adj}
\eeqa
The factor $(e_1-e_2)$ has been introduced so that $m$ is
proportional to the residue of $f_1(x,y)$ at $x=e_1$ with a
$\tau$-independent proportionality constant. Also, $u$ is related to the
Casimir of $SU(2)$ as can be seen in the limit $m=0$.

By defining the $Z_2$ invariant variables $t=vy$, $w=v^2$, we can use
(\ref{su2adj}) to solve for $w$ and subtitute into the Weierstrass
equation to obtain
\beqa
t^2\; =\; -(x-e_1)(x-e_3)(x-e_3) \big[ \,
\frac{m\, (e_1-e_2)}{(x-e_1)(x-e_2)} t + u \big]
\eeqa
Redefininig $t\to t-\frac 12 m (x-e_3)(e_1-e_2)$, we have the
equation
\beqa
t^2\; =\; (x-e_3) \;\big[ \;\frac 14 m^2(e_1-e_2)^2
(x-e_3)-u(x-e_1)(x-e_2)\;\big]
\eeqa
The discriminant of this curve contains a factor $\prod_{i<j}(e_i-e_j)^2$
corresponding to singularities in coupling space, and also shows the
existence of singularities in the Coulomb branch at $u=0$ and the zeroes
of $16 u^2 - 24 e_3 m^2 u + m^4 (e_1-e_2)^2$, which correspont to
those found in \cite{swsu2} \footnote{In order to do so, one must notice
that the gauge coupling of the $SU(2)$ is actually given by $\tau/2$ and
not $\tau$}.


That the curve (\ref{su2adj}) is correct could have been guessed from the
following argument. Notice that for $p=-1$ the multiplicity of zeroes
required do not impose new conditions. The curve (\ref{su2adj}) is similar
to the curve of a gauge theory $SU(2)\times SU(2)$ with two
bi-fundamentals \cite{witten4d}, but with an additional $Z_2$ symmetry.
The symmetry implies that only a sub-locus of the $SU(2)^2$ theory, and
only certain cycles in the curves are considered. Thus one is actually
describing the effective couplings of the diagonal $SU(2)$. It also has
charged matter in the adjoint and a singlet, which constitute the
symmetrized version of the bi-fundamentals.

This argument generalizes to all values of $n$. As long as $p=-1$, our
curves describe a symmetrized version of the $SU(n)^2$ theory with two
bi-fundamentals. The gauge group whose dynamics is determined is again
the diagonal $SU(n)$ subgroup. However, there are two possibilities for
its hypermultiplet content: either it is an adjoint and a singlet, or else
a symmetric and an antisymmetric. That the one realized is the second
follows from the two non-trivial flat directions studied above.

\bigskip
\medskip

We have described in some detail the construction of the curves for all
the possible gauge
theories which can be realized through brane configurations in the
presence of the $O6^+$-$O6^-$ orientifold background (in absence of
global mass). In all cases we have
constructed a spectral cover with the right number of parameters to
describe the dynamics, and we have reproduced the interesting connections
between them through flat directions. We have also directly compared our
answers with known results in some simple examples.

Let us mention that there are many other checks that are presumably
authomatically fulfilled by these curves, just because they are inherited
from analogous properties in Witten's elliptic models for unitary gauge
groups. For
example, it was shown in \cite{dw} that the curves for the $N=4$ $SU(n)$
theories gave the correct behaviour when one of the eigenvalues of the
adjoint scalar in the $N=2$ vector multiplet is much larger than the
others. The curve flows to that of the $SU(n-1)$ theory, and the brane
interpretation of the process is to send one D4-brane to infinity in $v$.
One can clearly think of similar flows in the general product group
theories, which must be reproduced by the curves in \cite{witten4d}. Since
we have been able to put the orientifolded theories in the very same
framework of elliptic models, it is clear that such flows will be
correctly reproduced by the curves we have constructed.

We think all of this provides strong evidence that the proposed curves
yield the correct description of the corresponding gauge theory dynamics.
In Section 6 we will explore the introduction of the global mass.
Before that, however, we would like to briefly discuss the realization of
Montonen-Olive duality in our framework.

\subsection{Montonen-Olive duality}

We have already stressed the importance of the observation of section 4.4
in matching the correct number of parameters lost along the flat
directions which connect our models. Another important consequence is the
interpretation of Montonen-Olive duality in our curves.

Since in M-theory the spacetime contains a torus, there are different
degenerations of it (shrinking different circles) related by
$SL(2,\IZ)$ that lead to possibly
different weakly coupled gauge theory interpretations.
Each such degeneration is roughly defined by a coalescence of
the four fixed points pairwise. In this respect, it is interesting to
abstractly classify the different limits that arise depending on how the
poles corrresponding to the NS5-branes at fixed points are located with
respect to the point $P_{O6^-}$, which defines the classical position of
the $O6^-$. We can then provide a gauge theory interpretation to the
resulting type IIA brane configuration.

{\bf a)} The first case is not having poles at fixed points. Any classical
limit of such configuration yields a theory in the family i) of models,
i.e. with no NS5-branes intersecting the O6-planes. In this sense, the
theories are self-dual under the $SL(2,\IZ)$ transformations.

{\bf b)} A second case consists on having a simple pole precisely at
$P_{O6^-}$. Again, any classical limit yields theories with a NS5-brane
intersecting the $O6^-$, i.e. in the family ii). Since the pole and the
zeroes of the elliptic functions are at the same fixed point, it follows
that the number of D4-branes must be even. For the particular case of
$p=-1$, any classical limit yields an $SO(2k)$ theory with matter in the
adjoint, and we have recovered the usual Montonen-Olive self-duality of
this theory.

{\bf c)} Another possibility is having a simple pole at a fixed point
different from $P_{O6^-}$. There are essentially two different classical
limits, related by S-duality. One of them is defined by the coalescence of
these two fixed points, and is interpreted in the type IIA language as a
model with a NS5-brane intersecting the $O6^-$. The gauge theory obtained
corresponds to the family ii) with an odd number of D4-branes. The other
limit is obtained by letting these two fixed point go to different
positions in the type IIA configuration. The NS5-branes then sits at the
$O6^+$ and the gauge theory described is of type iii). This is a
generalization of the Montonen-Olive duality between $SO(2k+1)$ and
$Sp(k)$ (recovered as a particular case for $p=-1$). The same curve can
reproduce a cover of type iii) with $2k+2p+2$ sheets and a model of type
ii) with $2k+2p+3$, because in the latter a whole sheet is frozen at
$v=0$.

{\bf d)} If there are two poles at fixed points, one of which precisely at
the position $P_{O6^-}$, there is a classical limit in which there is a
NS5-brane intersecting each of the O6-planes. The corresponding gauge
theories are of type iv) with an even number of D4-branes. The S-dual
limit does not
have a clear weak coupling interpretation, since it would correspond to
the coalescence of two poles, or the overlapping of the NS5-branes.

{\bf e)} Finally, there can be two poles at fixed points which both differ
from $P_{O6^-}$. The only limits for which a clear weak coupling
interpretation can be provided are those having one NS5-brane intersecting
each O6-plane. They correspond to theories in the family iv) with an odd
number of D4-branes.

\medskip

Any further addition of poles at fixed points yields theories which have
coincident NS5-branes when one shrinks any circle in the torus.

We have thus found a nice description of Montonen-Olive duality based on
the interplay of the positions of zeroes and poles in our curves. It is
also an interesting exercise to follow the flat directions emanating from
a given theory in two different languages. For example, starting with a
theory in the family iv) with an odd number of D4-branes (described in
e)), the removal of the very same pole can be interpreted as a Higgssing
to a model of type ii) (breaking $SU(2k+2p+3)$ to $SO(2k+2p+3)$ by means
of the symmetric representation) or as the complicated Higgssing
(\ref{comphig2}) to a model of type iii). Both flat directions are
exchanged by Montonen-Olive duality.

Let us stress that this description is based on the geometry of the torus.
Since this structure will also be present when we introduce the global
mass, it is reasonable to expect these dualities to hold even in that
case, as happens for elliptic models of the $SU(k)$ type \cite{dw,witten4d}.

\subsection{The $O6^-$-$O6^-$ background revisited}

To end this section, we briefly comment on the models obtained
in the presence of two negatively charged orientifold planes and four
D6-branes
(plus mirrors). Since the D6-branes are free to move in the $v$-direction,
as opposed to the O6-planes which must sit at the fixed points, the
sources for the RR charge are in this case distributed on the $v$-plane.
These positions are not hidden parameters and appear explicitly in the
curve. Therefore one has to deal with an ambient geometry which is not
rigid, but
has the structure of an elliptic fibrations over the space parametrized by
$v$. Some information about this fibration can be gained using the duality
with F-theory, as we shrink the fiber torus. For instance, for the
$Sp(k)$ theory with a {\em massless} antisymmetric and four flavours, this
space
is the Seiberg-Witten fibration for $SU(2)$ with four flavours
\cite{swsu2}. In the M-theory picture the spectral cover reduces to a set
of elliptic fibers over $k$-points in the base $u$-space (or mirror
pairs in the double cover parametrized by $v$). Upon shrinking the fiber,
the D4-branes become D3-brane probes of the non-trivial F-theory
background in which the type IIB $\tau$ parameter varies due to the
presence of the D7-branes and an O7-plane (of charge -8), coming from the
D6-branes and
the {\em two} ($O6^-$)-planes (note a similar comment in \cite{witfut} on
the
relation between O6- and O7-planes). Thus the picture of
\cite{senfth,bds,dls,aharony} is
recovered. An analogous though not completely identical T-duality has been
considered in \cite{johntdual}.

In principle this construction could be generalized to product group
theories, since the spacetime for M-theory is already known,
$E_{\tau}\times \IC$ is replaced by the SW fibration which gives a well
defined dependence of $\tau$ on $v$. One can also consider curves which
wrap a certain number of times around the generic elliptic fiber. These
generalize the spectral cover we introduced, and constitute
the SW curves for the theories. However, many technical issues should be
reconsidered in this more general framework. Clearly more work is
required to develop a further understanding of these models, even for
massless antisymmetric hypermultiplets.

\medskip

However, for massless flavours all the sources of RR charge are located at
$v=0$ and can be cancelled locally by moving the D6-branes only in the
$x^6$ direction. This allows for embedding the curves in a rigid
$E_{\tau}\times \IC_v$ spacetime, as we have done with the $O6^+$-$O6^-$
background. It is a simple exercise to derive that any movement
of D6-branes which leads to a local charge cancellation yields a well
defined cover for the curves of $\b=0$
theories. The simplest option is to cancel the charge of each $O6^-$ using the
charges of the two D6 closest to it \footnote{Other choices lead to
additional
sheets that, after one imposes the vanishing conditions, are frozen at
$v=0$ and do not influence the dynamics.}.

The configuration can be lifted to M-theory considering a cover of
$E_{\tau}$. The elliptic functions appearing in its equation must verify
the usual parity conditions. Since there are two $O6^-$'s, we will have to
impose vanishing conditions at both of their locations, the orders of the
zeroes depending on the number of NS5-branes the D6-branes must cross to
reach the orientifold plane. Again we cannot be completely explicit due to
the proliferation of cases in these models, but we hope the examples we
have considered in the $O6^+$-$O6^-$ configuration will be useful as a
guideline for future research in the $O6^-$-$O6^-$ background as well.

\section{Introduction of the global mass}

The construction of these gauge theories via the brane configurations
described in Section~3 allows the introduction of a non-vanishing sum of
hypermultiplet masses, i.e. this parameter is not projected out by the
orientifold action. This can be argued pictorically as in Figure~1, or by
noticing, as we did in subsection 5.1, that $f_1$ is an odd function under
the involution of the torus, and thus can have a simple pole at infinity,
which is the key ingredient for such a non-zero mass \cite{witten4d}.

It is not completely clear that the approach we have used remains valid if
a global mass is turned on. The viability of considering a trivial
spacetime relied on cancelling locally the RR charge by moving it along
$x^6$. It may be that the non-zero mass induces a relative shift of the
$v$-positions of the O6-planes, in which case we would have to deal with
a spacetime with the structure of a non-trivial elliptic fibration, much
in the way described in section 5.6.

It may however happen that the mass deformation can simply be taken into
account in analogy with the elliptic models with unitary gauge groups.
After all, this is the most natural deformation of the massless case, and
it is the simplest extension of the mechanism implemented in the $SU(n)$
case. We  have not found a complete description of all the families and
the flat directions connecting them. Let us however briefly report on a
possible construction of models of types i) and iii), hoping its further
analysis to help clarifying this important issue, which we leave for
future research.

\medskip

{\large {\em Family i)}}

The curve for this models with a global mass introduced is described, away
from infinity, by a $(2k+2p+2)$-fold cover of $E_{\tau}$ of the type (5.1).
The function $f_i(x,y)$ has the usual parity properties, and has simple poles
at the $2p+2$ mirror locations of the NS5-branes. It also has a pole of order
$i$ at infinity. This of course introduces new parameters in the equations, to
be determined by the conditions in a neighbourhood of the point at infinity.
If we locate $P_{O6^-}$ away from infinity, we must impose $f_{2k+l}(x,y)$
to have a zero of order $l$ at such point.

Upon the change of variables (\ref{patchmass}), with $k$ properly
substituted by $2k+2p+2$, the number of sheets in the cover, one obtains a
cover of the type (\ref{covertilde}). The functions ${\tilde f}_i$
verifies the parity properties, and in principle it has a pole of order
$i$ at infinity. However one must impose that the functions
${\tilde f}_1$, ${\tilde f}_2$, $\ldots, {\tilde f}_{2k}$ must actually be
holomorphic there (since there is no NS5-brane at that location).
It is easy to check that the counting of parameters matches the gauge
theory result, and that when the global mass vanishes, the conditions
reproduce the ones considered in Section~3.

As mentioned there, the $Sp(k)\times SO(2k+2)$ in this family has also
been constructed using an O4-plane \cite{lll}. A difference in both
approaches is that the O6-planes allow for a non-vanishing mass for the
bi-fundamental hypermultiplet, whereas the O4-plane does not. Notice that
such term is also possible from the field theory point of view (the
argument in \cite{lll} to explain its absence in other models does not
apply in this case), so our construction provides a more faithful
realization of the theory.

\bigskip

{\large {\em Family iii)}}

The introduction of global mass in this family requires a new ingredient.
There must be two additional sheets in the SW curve. The motivation for
introducing them will become clear in the consistency checks we are to
perform on the curve.

For definiteness let us locate the unpaired NS5-branes at the point at
infinity in $E_{\tau}$.We propose that the SW curve for the
theory with gauge group $Sp(k)$$\times SU(2k+2)$$\times \ldots$$\times
SU(2k+2p)$$\times SU(2k+2p+2)$ is a $(2k+2p+4)$-fold cover of $E_{\tau}$,
\beqa
v^{2k+2p+4} \, + \, f_1(x,y) v^{2k+2p+3}\, + \ldots \, +
\, f_{2k+2p+3}(x,y) v\, + \, f_{2k+2p+4}\; = \; 0.
\label{coverhuge}
\eeqa
The elliptic functions verify the usual parity properties. They have
simple poles at the locations of the $2p+2$ NS5-branes, and $f_i(x,y)$ has
a pole of order at most $i$ at infinity. Since the $O6^-$ is located away
from infinity, the conditions on the multiplicity of zeroes at its
position must be imposed to these functions. They imply that $f_{2k+l}$
has a zero of order $l$ at the $O6^-$.

Notice that in the case of
vanishing global mass, the poles at infinity are not present (but for a
simple one) and the vanishing conditions on $f_{2k+2p+3}$ and
$f_{2k+2p+4}$ are strong enough to force them to vanish identically.
Thus the two additional sheets in (\ref{coverhuge}) are frozen at $v=0$
and do not enter the dynamics, and the description in Section~3 in terms
of a $(2k+2p+2)$-fold cover is correct.

Finally, one must impose that the functions ${\tilde f}_i$, obtained
through a change to good variables in a neighbourhood of the point at
infinity, must have at most a simple pole at this point, since there is a
NS5-brane there. It is easy to check that the counting of parameters gives
the correct answer, matching the gauge theory result.

\medskip

The main argument for increasing the number of sheets in the cover by two
and precisely two units comes from obtaining a good flow of the mass
deformed $N=4$ $Sp(k)$ theory to the pure $Sp(k)$ one. This is done
following \cite{dw} (a geometric argument analogous to the one in
\cite{donagi} is also possible \cite{don1}).

When the only NS5-brane (recall that $p=-1$) is located at infinity, the
$(2k+2)$-fold cover can be written in terms of some polynomials
$Q_{2l}(v,x,y)$, of degree $2l$ in $v$. They satisfy the properties of the
polynomials $P_i(v,x,y)$ of reference \cite{dw} plus the additional
condition that $Q_{2l}(v=0,x,y)$ has a double zero at $P_{O6^-}$, say
$x=e_1$. For each $l$ there is a unique such polynomial.
In terms of them, the $(2k+2)$-fold cover is
 \beqa
Q_{2k+2}(v,x,y)\, + A_2 Q_{2k}(v,x,y) \, + \ldots \,+ A_{2k} Q_{2}(v,x,y)
\; = \; 0.
\label{coversp}
\eeqa
(notice that $Q_0(v,x,y)$ vanishes identically). It is easy to compute the
first polynomials, even though they soon become lengthy. For definiteness,
we illustrate the $N=2$ flow for $Sp(1)$, and then comment on the general
pattern for higher ranks. The cover for $k=1$ is given by
\beqa
\big[ \; t^4-6(x-e_1)t^2 & + & 8yt-3(x-e_1)^2+4(x-e_1)\sum_{i=1}^3 e_i -12
e_1(x-e_1)\; \big] + \nonumber \\
& + & A_2\;[\,t^2-x+e_1\,] \; =\; 0
\label{sp1}
\eeqa
Before taking the limit to the pure gauge theory it is convenient to
perform an affine transformation in $x$ so that the Weierstrass equation
reads
\beqa
y^2\;=\; x(x-\lambda)(x-1)
\label{waffine}
\eeqa
The order parameter $A_2$ suffers an additive renormalization and will be
denoted $A_2^{\prime}$ in what follows.

The desired limit can be taken by letting all dimensional parameters go to
zero as an adequate power of $\lambda\approx q^{1/2}$. The renormalization
group matching condition for $Sp(k)$ is $\Lambda^{4k+4}=q\, m^{4k+4}$
(with $\Lambda$ being the dynamical scale of the pure gauge theory), so the
appropriate scalings are
\beqa
& v\to \lambda^{\frac{1}{2k+2}}{\tilde v} \quad ; \quad & A_i^{\prime} \to
\lambda^{\frac{i}{2k+2}} a_i \nonumber\\
& x\to \lambda {\tilde x} \quad ; \quad & y\to \lambda {\tilde y}
\eeqa
with ${\tilde v}$, $a_i$, ${\tilde x}$, and ${\tilde y}$ finite in the
limit $\lambda\to 0$.

Also, if we are to keep the $O6^-$ at finite
distance, we should choose $e_1$ to be either the $x=0$ or $x=\lambda$
roots in (\ref{waffine}). As $\lambda\to 0$, this equation reduces to
\beq
{\tilde y}^2 \; = \; -({\tilde x} - 1/2)^2 \, + \, \frac{1}{4}.
\label{weierslim}
\eeq
which describes $\IR\times S^1$. The two choices for $e_1$ map to
${\tilde x}=0,1$, the two fixed points in the non-compact $x^6$ limit.

For $k=2$, the dominant terms in the equation (\ref{sp1}) are those
scaling with a single power of $\lambda$. They yield the equation
\beqa
{\tilde t}^4\, +\, 4 {\tilde x}\, +\, a_2 {\tilde t}^2\;=\; 0
\eeqa
if $e_1=0$, or the same equation with ${\tilde x}-1$ in place of
${\tilde x}$ if $e_1=\lambda$. Solving for ${\tilde x}$ and substituting
in (\ref{weierslim}), we obtain
\beqa
{\tilde y}^2 \; = \; -\big[\;{\tilde t}^2({\tilde t}^2+a_2) \pm
\frac{1}{2}\;\big]^2 \, + \, \frac{1}{4}.
\eeqa
which is precisely the curve for the pure $Sp(1)$ theory as expressed in
\cite{dhkp}. The two signs correspond to the two choices of $P_{O6^-}$
mentioned above.

The pattern in this flow is identical for other values of $k$. The
existence of the two additional sheets ensures a scaling appropriate for
the RG matching conditions for $Sp(k)$. Also, there is always a term
proportional to $(x-e_1)$ in $Q_{2k+2}$ which allows to solve for
${\tilde x}$. This last has the structure ${\tilde x}={\tilde v}^2
p_k({\tilde v}^2)$,
with $p_k({\tilde v}^2)$ being the characteristic polynomial containing
the order parameters of $Sp(k)$. The absence of a constant term in the
right hand side of this equation follows from the condition of the zero in
$Q_{2l}(v,x,y)$. Upon substitution in (\ref{weierslim}) the curve
in \cite{dhkp} is recovered.

Since these curves should also describe the $SO(2k+1)$ theories with
a massive adjoint hypermultiplet, there should be a flow to
the curve for the pure $SO(2k+1)$ theory. This issue is under current
investigation and would provide a nice check of our proposal.

A general argument favouring the additional sheets in configurations with
more NS5-branes is also possible. In the gauge theory $Sp(k)$$\times
SU(2k+2)$$\times \ldots$$\times SU(2k+2p)$$\times SU(2k+2p+2)$, with
bi-fundamentals, and an antisymmetric for the last factor,
the decoupling of this last hypermultiplet can be performed by increasing
the global mass and keeping the bi-fundamental masses constant. The theory
after decoupling the antisymmetric is reproduced by a model with
non-compact $x^6$. This flow can be considered in close analogy with a
similar flow in a $SU(2k+2p+2)^{p+2}$ Witten elliptic model, in which
$Z_2$ invariance is imposed, and some parameters are frozen by vanishing
conditions \footnote{I would like to thank K.~Landsteiner for bringing to
my attention the analogy of this argument with the viewpoint of
ref.\cite{dhkp}.}. The orders of vanishing of the functions in the final
curve after the flow are only reproduced if one starts with a
$(2k+2p+4)$-fold cover with the vanishing conditions stated as above.

\medskip

Let us finish mentioning how the Higgssing to models of type i) would proceed.
Since the number of sheets drops by two, in the brane picture it
corresponds to removing the NS5-brane at $O6^+$ along with two D4-branes
attached to it. In the equations this is accomplished by the the
factorization of the cover (\ref{coverhuge}) in two pieces when $k+p+1$
paremeters are appropriately tuned. One of the pieces describes a cover of
type i) with a global mass, whereas the other describes the branes we have
removed from the main brane configuration.

\section{Conclusions}

We have introduced the brane configurations appropriate for the study of
the mass deformed $N=4$ theories with symplectic and orthogonal gauge
groups. The solution of these models would follow by embedding these
configurations in M-theory. The complicated geometries associated to the
orientifold planes can be avoided when the sum of the hypermultiplet
masses vanishes, and we have shown the construction of the curves in close
analogy with the elliptic models with unitary gauge groups.

Unfortunately, the most interesting cases, namely symplectic or orthogonal
groups with adjoint matter, are too trivial if this mass vanishes. Even
though we have made a partial proposal based on a rigid M-theory
spacetime, the solution of the mass deformed models may require the full
orientifold geometry to come into play. The clarification of whether this
is so is the next issue to be addressed for the understanding of these
theories.

Finally, we have determined the geometric features on the curves that
explain Montonen-Olive duality in the  whole class of $N=2$ models
studied. In
particular, the exchange of $Sp(k)$ and $SO(2k+1)$ groups, and the
self-dualtiy
of $SO(2k)$ receive a simple interpretation in terms of how the fixed
points of the $Z_2$ action coalesce in the different classical limits.
This mechanism is expected to hold even when the global mass is turned on.

\bigskip

\centerline{\bf Acknowledgments}

I would like to thank G.~Curio, A.~Hanany, D.~A.~Lowe, J.~H.~Schwarz and
E.~Witten for useful comments, R.~Donagi for stimulating correspondence,
and M.~Gonz\'alez and L.~E.~Ib\'a\~nez for advice and
encouragement. This work is supported by a Ram\'on Areces
Foundation fellowship.

\end{document}